\documentclass[usenatbib]{mn2e}

\usepackage{graphicx} 
\usepackage[applemac]{inputenc}
\usepackage{epsfig}
 \usepackage{comment}
\usepackage{natbib}
 \usepackage{lscape}
 \usepackage{colortbl}
 \usepackage{verbatim} 
\usepackage{url}
 \bibpunct{(}{)}{;}{a}{}{,}
\usepackage{amsmath}
\usepackage{amssymb}
\usepackage{caption}
\usepackage[caption=false]{subfig}
\usepackage{tabularx}
\usepackage{booktabs}

\title[MSR in different environments]{Larger sizes of massive quiescent early-type galaxies in clusters than in the field at $0.8<z<1.5$ }
\author[L. Delaye et al.]
  {L.~Delaye$^{1,11}$,
  M.~Huertas-Company$^{1,2}$, S.~Mei$^{1,2}$, C.~Lidman$^3$, R.~Licitra$^1$, A.~Newman$^4$,   
  \newauthor
A.~Raichoor$^1$, F.~Shankar$^1$, F. Barrientos$^5$, M.~Bernardi$^6$, P.~Cerulo$^7$,  W.~Couch$^7$, 
\newauthor
R.~Demarco$^8$, R. Mu\~noz$^5$,R.~S\'anchez-Janssen$^9$, M.~Tanaka$^{10}$ \\
$^{1}$GEPI, Paris Observatory, 77 av. Denfert Rochereau, 75014 Paris, France\\
$^{2}$University Denis Diderot, 4 Rue Thomas Mann, 75205 Paris, France\\
$^{3}$Australian Astronomical Observatory, PO Box 296 Epping, NSW 1710 Australia\\
$^{4}$Cahill Center for Astronomy and Astrophysics, California Institute of Technology, MS 249-17, Pasadena, CA 91125, USA\\
$^{5}$Department of Astronomy and Astrophysics, Universidad Cat\'olica de Chile, Santiago, Chile\\
$^{6}$Department of Physics and Astronomy, University of Pennsylvania, 209 South 33rd Street, Philadelphia, PA 19104, USA\\
$^{7}$Centre for Astrophysics \& Supercomputing, Swinburne University of Technology, PO Box 218, Hawthorn, VIC 3122, Australia\\
$^{8}$Department of Astronomy, Universidad de Concepción, Casilla 160-C, Concepción, Chile\\
$^{9}$European Southern Observatory, Alonso de Córdova 3107 Vitacura, Santiago Chile\\
$^{10}$Institute for the Physics and Mathematics of the Universe, The University of Tokyo, 5-1-5 Kashiwanoha, Kashiwa-shi, Chiba 277-8583 Japan\\
$^{11}$CEA-Saclay, DSM/IRFU/SAp, F-91191 Gif-sur-Yvette, France}

\begin{document}
\date{Released 2002 Xxxxx XX}

\pagerange{\pageref{firstpage}--\pageref{lastpage}} \pubyear{2002}

\maketitle

\label{firstpage}

\begin{abstract}
	The mass-size relation of early-type galaxies has been largely studied in the last years to probe the mass assembly of the most massive objects in the Universe. Recent work has established that their median size at fixed stellar mass increased by a factor of a few in the last $\sim10$~Gyrs. The exact mechanisms provoking this growth are still a matter of debate. We have shown in previous work how the study of the mass-size relation in different environments can provide important constraints to galaxy evolution models. 
	
	In this paper, we focus on the mass-size relation of quiescent massive ETGs ($M_*/M_\odot>3\times10^{10}$) living in massive clusters ($M_{200}\sim10^{14} \; \text{M}_\odot$) at $0.8<z<1.5$, as compared to those living in the field at the same epoch. Our sample contains $\sim 400$ ETGs in clusters and the same number in the field. Therefore, our sample is approximately an order of magnitude larger than previous studies in the same redshift range for galaxy clusters.

We find that ETGs living in clusters are between $\sim30-50\%$ larger than galaxies with the same stellar mass residing in the field. We parametrize the size using the mass-normalized size, $\gamma=R_e/M_{11}^{0.57}$. The $\gamma$ distributions in both environments peak at the same position but the distributions in clusters are more skewed towards larger sizes. Since this size difference is not observed in the local Universe, the size evolution at fixed stellar mass from $z\sim1.5$ of cluster galaxies is less steep ($\propto(1+z)^{-0.53\pm0.04}$) than the evolution of field galaxies ($\propto(1+z)^{-0.92\pm0.04}$). The size difference seems to be essentially driven by the galaxies residing in the clusters cores ($R<0.5\times R_{200}$). If part of the size evolution is due to mergers, the difference we see between cluster and field galaxies could be due to higher merger rates in clusters at higher redshift, probably during the formation phase of the clusters when velocity dispersions are lower. We cannot exclude however that the difference is driven by newly quenched galaxies which are quenched more efficiently in clusters. The implications of these results for the hierarchical growth of ETGs will be discussed in a companion paper.


\end{abstract}


\section{Introduction}

The mass assembly of the most massive galaxies in the universe is still an open issue. For a long time, the uniformity of their stellar populations together with their regular morphology have been interpreted as signs of a relatively quiet life, dominated by a strong starburst at very early epochs followed by a passive evolution \citep{partridge67,larson75}. 
The discovery of a population of massive early-type galaxies (ETGs) at high redshift, {\bf on average} $2-5$ times more compact than their local counterparts is the evidence of a size growth that is not predicted by this simple scenario \citep{daddi05, trujillo06}. Even though the exact abundance of these compact objects in the local universe is still debated today \citep{valentinuzzi10, poggianti12, trujillo12}, it is accepted that at least a fraction of massive ETGs needs to significantly increase the size over the last $10$~Gyrs \citep[][and references therein]{buitrago08, vandokkum08, saracco11, martinez11, vandesande11, raichoor12, newman12}. 

Two physical processes are usually invoked to explain such a growth but none of them is able to reproduce all the observed trends. \\
Intense AGN activity can expel the gas of the galaxy in a relatively short amount of time leading to a redistribution of the gravitational potential and hence to an increase of the size \citep{fan08,fan10}. Recent numerical simulations by \cite{ragone11} have shown however that the typical time scale for this process is only of a few Myrs which seems difficult to reconcile with the low dispersion in the ages and sizes of compact galaxies at high redshift (e.g. Trujillo et al. 2011). \\
On cosmological time scales, dry minor mergers can also lead to a growth of the galaxy by spreading stars in the outer parts after the merger event, without significantly increasing the stellar mass \citep{naab09, bezanson09, hopkins09a, bernardi09, shankar10b, shankar11, vandokkum10}. This scenario is particularly attractive because it can explain many of the observed properties of these objects (scatter, inside out growth...) and minor merging are known to be frequent events in a CDM cosmology. Extremely deep imaging of nearby ETGs has indeed revealed signs of disturbances in the outskirts of many of these galaxies \citep{duc11}. The direct observation of such minor mergers remains a challenge at high redshift, and there are still several open questions (e.g. \citealp{2013MNRAS.tmp.1437D}). \cite{newman12} found that the number of observed satellites around massive galaxies can account for the measured size growth from $z=1$ only if short dynamical time scales are assumed. Similar conclusions are also reached by \cite{lopezsanjuan12} who invoked a progenitor bias to explain the excess of growth they measure. \cite{huertas12} also showed that several hierarchical models based on the Millennium merger trees struggle to fully reproduce the amount of evolution reported by the data at fixed stellar mass.

Environment is an additional variable that can be analyzed to disentangle between different scenarios, as we have shown in several recent works (Raichoor et al. 2012, Huertas-Company et al. 2013, Shankar et al. 2013). Several hierarchical models predict indeed a correlation between galaxy size and the environment in which the galaxy lives, with larger galaxies in denser environments \citep[][ in preparation]{shankar11	}. Observational results up to now have been controversial though. In the local universe, Huertas-Company et al. (2013) did not find any trend with environment (see also Guo et al. 2010, Weinmann et al. 2009) for massive galaxies in the SDSS while \cite{poggianti12} found that cluster galaxies are slightly smaller than field galaxies at fixed stellar mass (see also Valtentinuzzi et al. 2010). Both results are at variance with first level predictions of some semi-analytical models (Huertas-Company et al. 2013). At $z<1$, \cite{huertas12} again did not find any difference between group and field galaxies at $z<1$ whearas \cite{cooper12} found that galaxies living in denser environments are larger although they measured the density of the environment in a different way. At higher redshifts ($z>1$) the situation is even worse since having a statistically significant sample of massive clusters at $z>1$ was almost impossible until very recently. The first works exploring that redshift range were based on one single cluster \citep{raichoor12, papovich12} and results are not in agreement. 

In this work, we make a step forward by analyzing a sample of 9 well known massive clusters ($M_h \sim 10^{14} \; \text{M}_{\odot}$) at $0.8 \lesssim z \lesssim1.5$ from the HAWK-I Cluster survey (Lidman et al., in preparation), to look for differences in the sizes of massive ETGs in cluster and field environments. All clusters but two have extended X-ray emission, between $20$ and $100$ spectroscopically confirmed members and have been observed with at least two filters with the Hubble Space Telescope (HST) Advanced Camera for Surveys (ACS).

The paper is organized as follows: in section 2, we present the dataset and describe the general methodology used to estimate sizes, masses and morphologies of cluster galaxies. In section 3, we describe the field galaxy sample used for comparison. We show our results in section 4 and discuss them in section 5. 

Throughout the paper, magnitudes are given in the AB system  (Oke \& Gunn 1983; Sirianni et al. 2005)  for all passbands. We assume a standard cosmological model with $\Omega_{M} = 0.3$, $\Omega_{\Lambda} = 0.7$ and $H_0 = 70 \; \text{km} \; \text{s}^{-1} \; \text{Mpc}^{-1}$ {\bf and use a Chabrier IMF}.

\section{Data and sample selection}
\label{sec:clusters}

\subsection{Cluster selection}

Our targets have been selected  according to the following criteria: (1) they cover a broad redshift range $0.84<z<1.45$; (2) they have been imaged with the HST/ACS in at least two bandpasses and have deep ground-based images in the near-IR; (3) they have at least 10 spectroscopically confirmed cluster members. 

All clusters have HST/ACS WFC (Wide Field Camera) images in at least two bandpasses. The ACS WFC resolution is 0.05~\arcsec/pixel, and its  field of view is 210~\arcsec x 204~\arcsec. The ACS/WFC PSF width is around $0.11 \arcsec$.
Our ACS/WFC images  were mostly obtained in a program designed to find Type Ia supernovae in distant galaxy clusters \citep{dawson09}. See \citet{meyers12} for a description of how these data were processed. Three clusters (see below): RDCS~J1252-2927, XMMU~J2235.3-2557 and RX~J0152-1357  had been previously targeted with the ACS camera on HST in the context of the ACS Intermediate Redshift Cluster Survey \citep{ford04, postman05, mei09} and these data have
been included. 


Eight of the nine clusters in this paper were targeted in the European Southern Observatory (ESO) HAWK-I\footnote{High Acuity Wide-field K-band Imager} cluster survey (HCS: Lidman et al. in prep). The HCS is a near-IR imaging survey that targeted nine well known high redshift galaxy clusters between $z=0.8$ and $1.5$. The aim of the survey was to obtain deep, high-resolution images of a sample of clusters for the purpose of studying the impact of environment on the evolution of cluster members. The ninth cluster in our sample, RDCS~J1252-2827, was imaged with ISAAC\footnote{Infrared Spectrometer And Array Camera} \citep{lidman04}. For some clusters, we also use J-band images from ESO/SofI\footnote{Son of ISAAC}.  

\begin{figure}
\includegraphics[width = 0.48\textwidth]{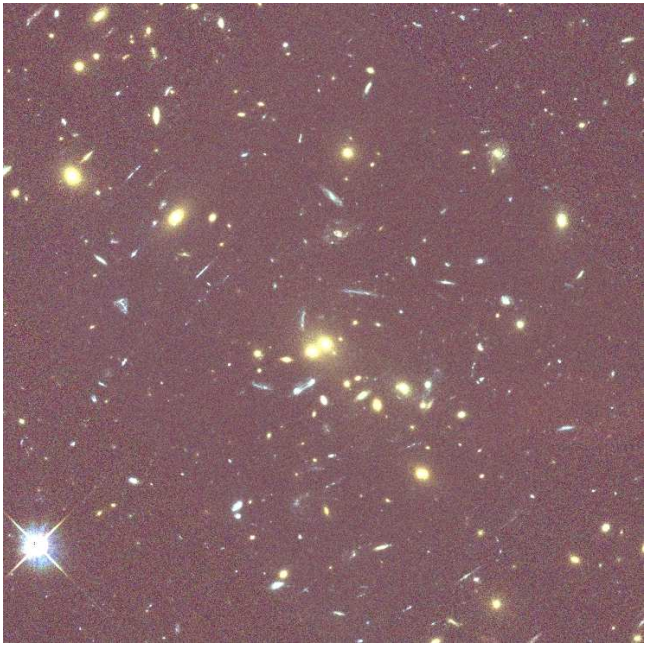}
\label{RX0152}
\caption{Color image of the centre of RX~J0152-1357 with $r_{625}$, $i_{775}$ and $z_{850}$ bandpasses. The field size is $75\times75$~arcsec, corresponding to $572\times 572$~kpc at $z=0.84$.}
\label{fig:cluster1}
\end{figure}

\begin{figure}
\includegraphics[width = 0.48\textwidth]{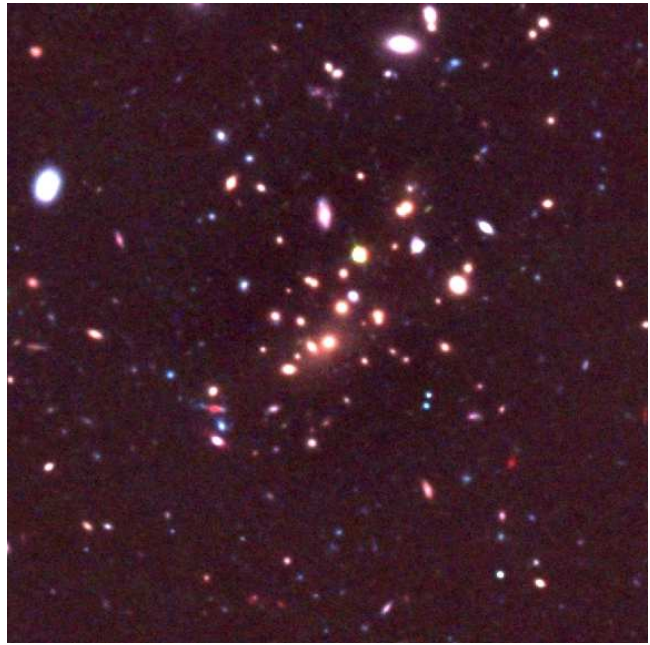}
\label{XMM1229}
\caption{Color image of the centre of XMMU~J1229+0151 with $i_{775}$, $z_{850}$ and $K_s$ bandpasses. The field size is $72 \times 72$~arcsec, corresponding to the same physical size as fig.~\ref{fig:cluster1}.}	
\label{fig:cluster2}
\end{figure}

 A summary of the observations is given in Table~\ref{tabobs} and the physical properties of each cluster are summarized in table~\ref{tabprop} (see also appendix~\ref{sec:append_cl} for more details on each individual cluster). We also show in figures~\ref{fig:cluster1} and~\ref{fig:cluster2} a color images of two of the clusters in our sample.

\begin{table*}
\begin{center}
    \begin{tabularx}{\textwidth}{p{5cm} X p{1cm}}
    \hline
	\hline
	\toprule
\multicolumn{1}{>{\raggedright}X}{Cluster} & \multicolumn{1}{>{\raggedright}X}{Filters} & \multicolumn{1}{>{\raggedright}X}{$N_{z_{spec}}$}\\
\midrule
\hline
RX~J0152-1357    &   $r_{625}$ ($19000$), $i_{775}$ ($19200$), $z_{850}$ ($19000$), Ks$^1$ ($9600$) & 107$^a$  \\
RCS~2319+0038    &   $i_{775}$ ($2400$), $z_{850}$ ($6800$),  Ks$^1$ ($9600$), J$^2$ ($2970$)       & 28$^b$   \\
XMM~J1229+0151   &   $i_{775}$ ($4110$), $z_{850}$ ($10940$), Ks$^1$ ($11310$), J$^3$ ($2280$)     & 26$^c$   \\
RCS~0220-0333    &   $i_{775}$ ($2955$), $z_{850}$ ($14420$), Ks$^1$ ($9600$), J$^2$ ($3330$)       & 11$^d$   \\
RCS~2345-3633    &   $i_{775}$ ($4450$), $z_{850}$ ($9680$),  Ks$^1$ ($9600$), J$^2$ ($2520$)       & 23$^d$   \\
XMM~J0223-0436   &   $i_{775}$ ($3380$), $z_{850}$ ($14020$), Ks$^1$ ($9600$), J$^1$ ($11040$)       & 27$^e$   \\
RDCS~J1252-2927  &   $i_{775}$ ($29945$), $z_{850}$ ($57070$), Ks$^2$ ($81990$), J$^2$ ($86640$)      & 31$^f$   \\
XMMU~2235-2557   &   $i_{775}$ ($8150$), $z_{850}$ ($14400$), Ks$^1$ ($10560$), J$^1$ ($10740$)     & 34$^g$   \\
XMM~J2215-1738   &   $i_{775}$ ($3320$), $z_{850}$ ($16935$), Ks$^1$ ($9600$), J$^1$ ($14400$)       & 48$^h$   \\
    \hline
  \end{tabularx}
  \caption{HAWK-I Cluster Survey data. $^1$ from HAWK-I, $^2$ from ISAAC, $^3$ from SOFI. $^a$\protect\cite{demarco05, demarco10}, $^b$\protect\cite{gilbank08, gilbank11, meyers10}, $^c$\protect\cite{santos09}, $^d$\protect\cite{meyers10,gilbank11}, $^e$\protect\cite{bremer06,meyers10}, $^f$\protect\cite{demarco07}, $^g$\protect\cite{rosati09}, $^h$\protect\cite{hilton10}.}
\label{tabobs}
\end{center}
\end{table*}

\begin{table*}
\centering
   \begin{tabular}{ c  c  c  c  c  c  c }
   \hline
	\hline
	Cluster & $z_{cl}$ & $\sigma_{vel}$ & $T$ & $M_{200}^X$  & $R_{200}$  & 
$M_{200}^L$\\
    &          & (km/s) & (keV) & ($10^{14} \; \text{M}_{\odot}$) & (Mpc)
& ($10^{14} \; \text{M}_{\odot}$)  \\
\hline
RXJ0152-1357    & $0.84$ &  $919 \pm 168$$^a$ &  $6.7 \pm 1.0$$^1$      & 
$7.3^{+1.8}_{-1.7}$  &  $1.17^{+0.09}_{-0.06}$ &  $4.4^{+0.7}_{-0.5}$ \\
RCS2319+0038    & $0.90$ &  $1202 \pm 233$$^b$ &  $6.2^{+0.9}_{-0.8}$$^2$ &
$5.4^{+1.2}_{-1.0}$  &  $1.22^{+0.15}_{-0.13}$ &  $5.8^{+2.3}_{-1.6}$ \\
XMMJ1229+0151   & $0.98$ &  $683 \pm 62$$^c$  &  $6.4^{+0.7}_{-0.6}$$^3$ &
$5.7^{+1.0}_{-0.8}$  &  $1.12^{+0.11}_{-0.10}$ &  $5.3^{+1.7}_{-1.2}$ \\
RCS0220-0333    & $1.03$ &  ...     &    ...        &   ...         & 
$1.09^{+0.12}_{-0.11}$   &  $4.8^{+1.8}_{-1.3}$ \\
RCS2345-3633    & $1.04$ &  $670 \pm 190$$^d$ & ...    &     ...        & 
$0.87^{+0.11}_{-0.10}$   &  $2.4^{+1.1}_{-0.7}$ \\
XMMJ0223-0436   & $1.22$ &  $799 \pm 129$$^e$ &  $3.8^{...}_{-1.9}$$^4$ & 
$2.4^{...}_{-1.5}$  &  $1.18^{+0.12}_{-0.11}$ &  $7.4^{+2.5}_{-1.8}$ \\
RDCSJ1252-2927  & $1.23$ &  $747^{+74}_{-84}$$^f$ &  $7.6 \pm 1.2$$^5$  & 
$4.4^{+1.1}_{-1.0}$  &  $1.14^{+0.06}_{-0.06}$ &  $6.8^{+1.2}_{-1.0}$ \\
XMMU2235-2557   & $1.39$ &  $802^{+77}_{-48}$$^g$ & 
$8.6^{+1.3}_{-1.2}$$^6$ &  $6.1^{+1.4}_{-1.2}$  &  $1.13^{+0.08}_{-0.07}$
&  $7.3^{+1.7}_{-1.4}$ \\
XMMJ2215-1738   & $1.45$ &  $720 \pm 110$$^h$ &  $4.1^{+0.6}_{-0.9}$$^7$ &
$2.0^{+0.5}_{-0.6}$  &  $0.9^{+0.17}_{-0.14}$ &  $4.3^{+3.0}_{-1.7}$ \\
   \hline
 \end{tabular}
 \caption{Cluster physical properties from \protect\cite{jee11}. {\bf From left to right, columns show the cluster redshift, the velocity dispersion, the X-ray temperature, the X-ray mass, the virial radius and the lensing mass.}
$^a$\protect\cite{demarco05}, $^b$\protect\cite{faloon13}, $^c$\protect\cite{santos09}, $^d$\protect\cite{jee11}, $^e$\protect\cite{meyers12}, $^f$\protect\cite{demarco07}, $^g$\protect\cite{rosati09}, $^h$\protect\cite{hilton10}, $^1$\protect\cite{ettori09}, $^2$\protect\cite{hicks08}, $^3$\protect\cite{santos09}, $^4$\protect\cite{bremer06}, $^5$\protect\cite{ettori09}, $^6$\protect\cite{rosati09}, $^7$\protect\cite{hilton10}.}
\label{tabprop}
\end{table*}

\subsection{Photometry and object detection}
\label{sec:photo}

Object detection is performed in the Ks band using Sextractor \citep{bertin96} and then the resulting catalog is matched in the other wavelengths, to obtain \texttt{MAG\_AUTO} magnitudes. Colors are computed with aperture magnitudes within an effective radius for each galaxy (estimated from the 2D Sersic best fit - see section~\ref{sec:sizes}), to avoid systematics due to internal galaxy gradients \citep{vandokkum98, vandokkum00, scodeggio01}. 

\begin{figure*}
  \centering
  \begin{tabular}{ c c }
\subfloat[]{\includegraphics[width = .49\textwidth, trim = 0.8cm 0.4cm 0.8cm 0.5cm, clip]{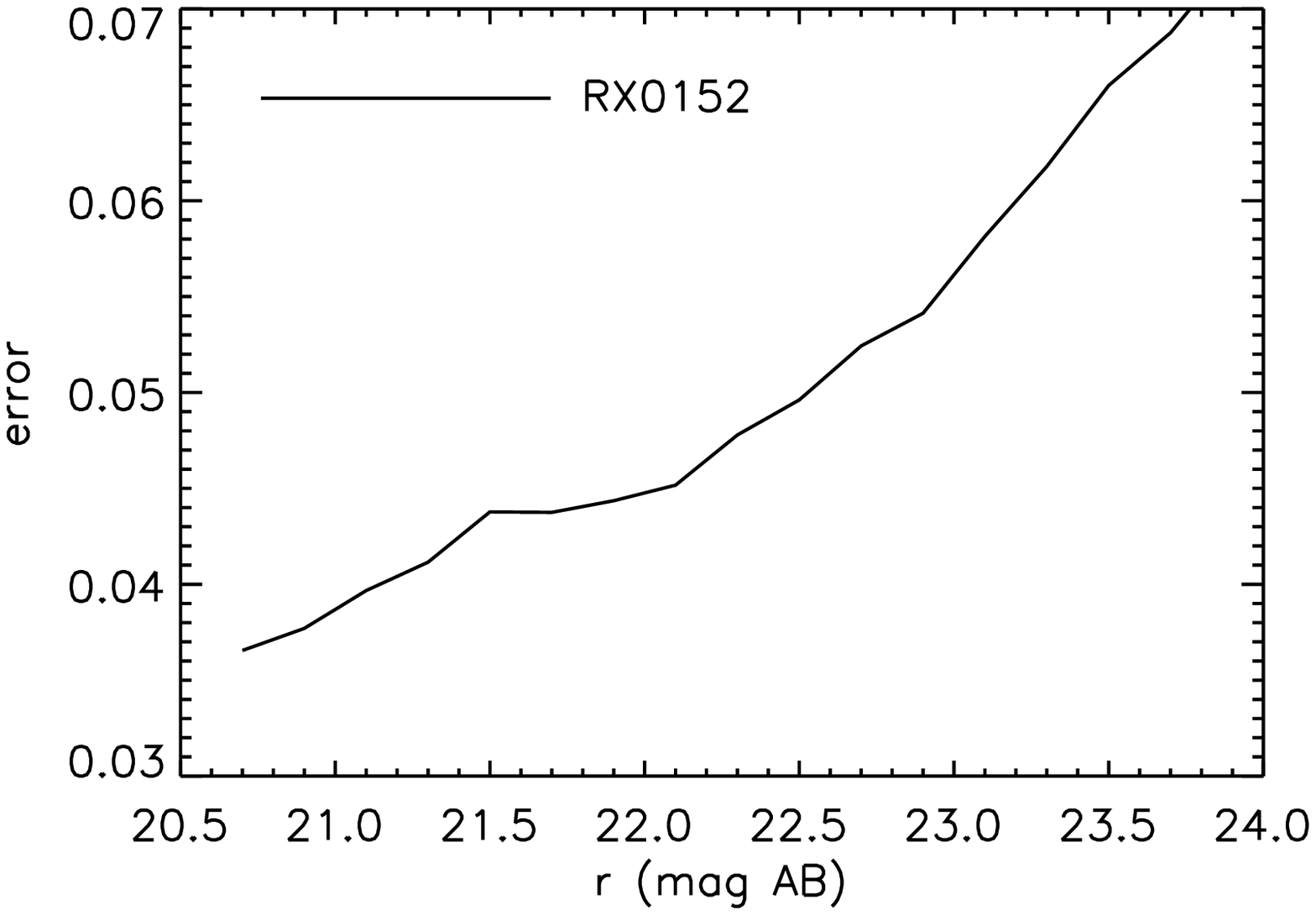}} & \subfloat[]{\includegraphics[width = .49\textwidth, trim = 0.8cm 0.4cm 0.8cm 0.5cm, clip]{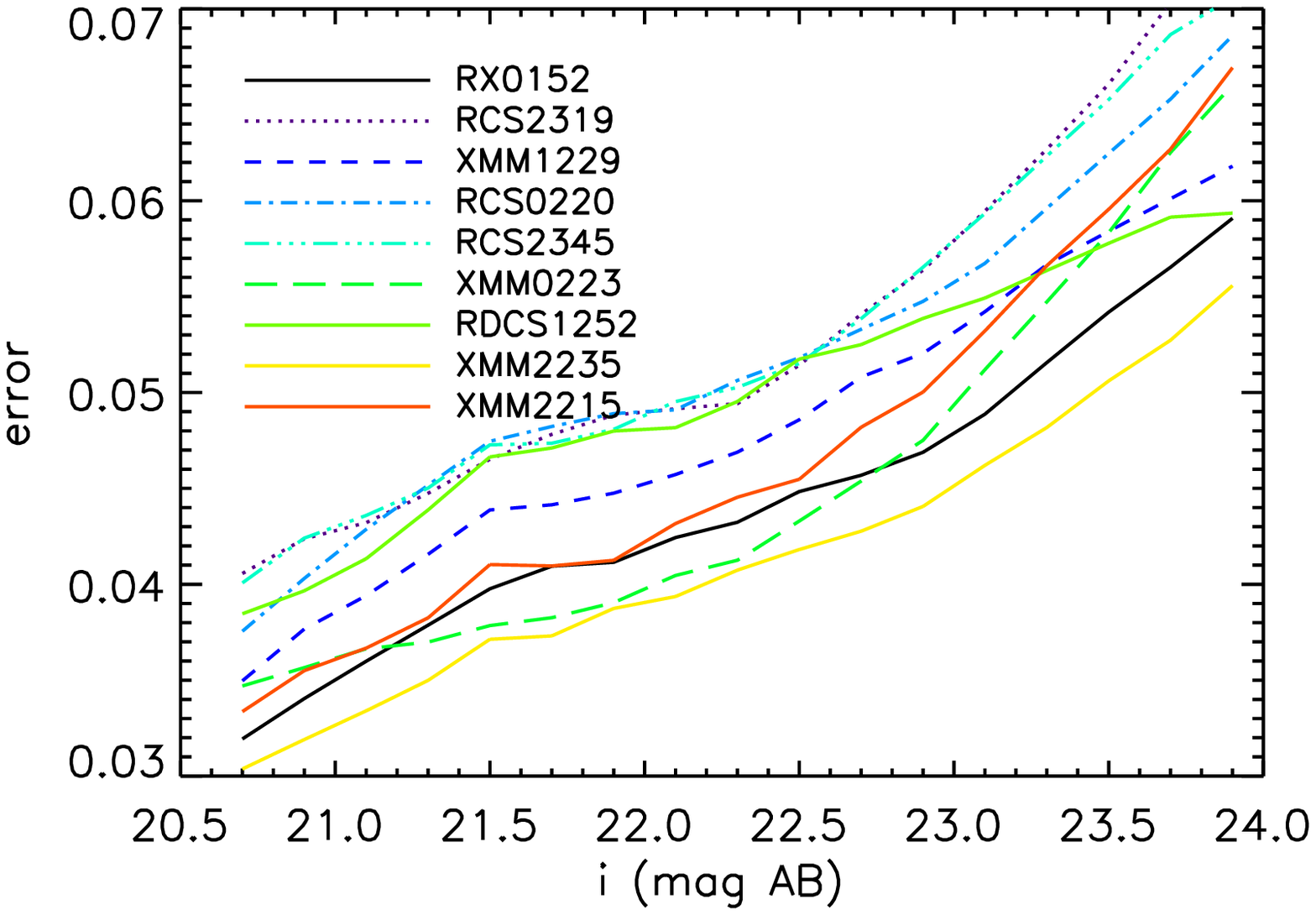}} \\
\subfloat[]{\includegraphics[width = .49\textwidth, trim = 0.8cm 0.4cm 0.8cm 0.5cm, clip]{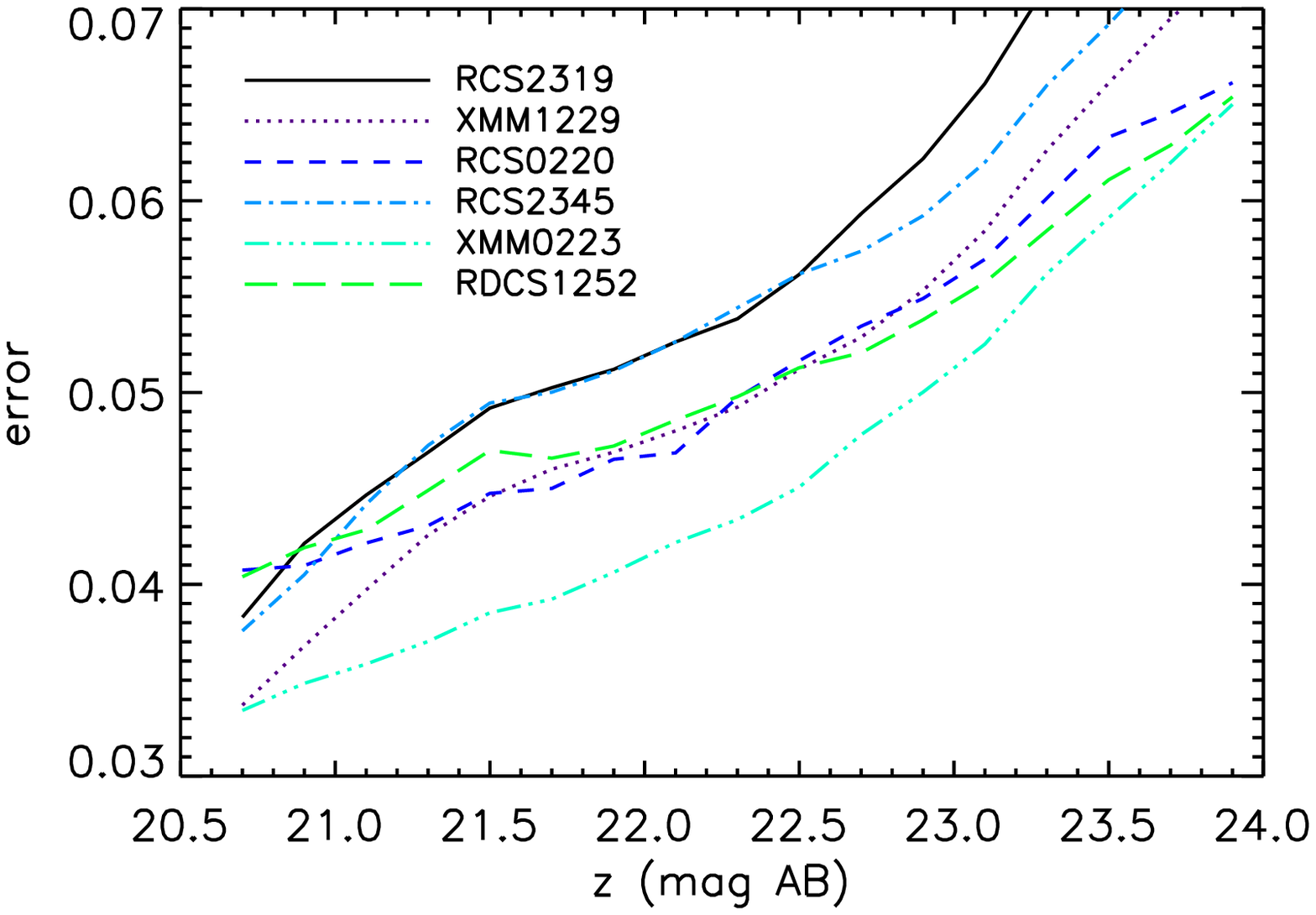}} & \subfloat[]{\includegraphics[width = .49\textwidth, trim = 0.8cm 0.4cm 0.8cm 0.5cm, clip]{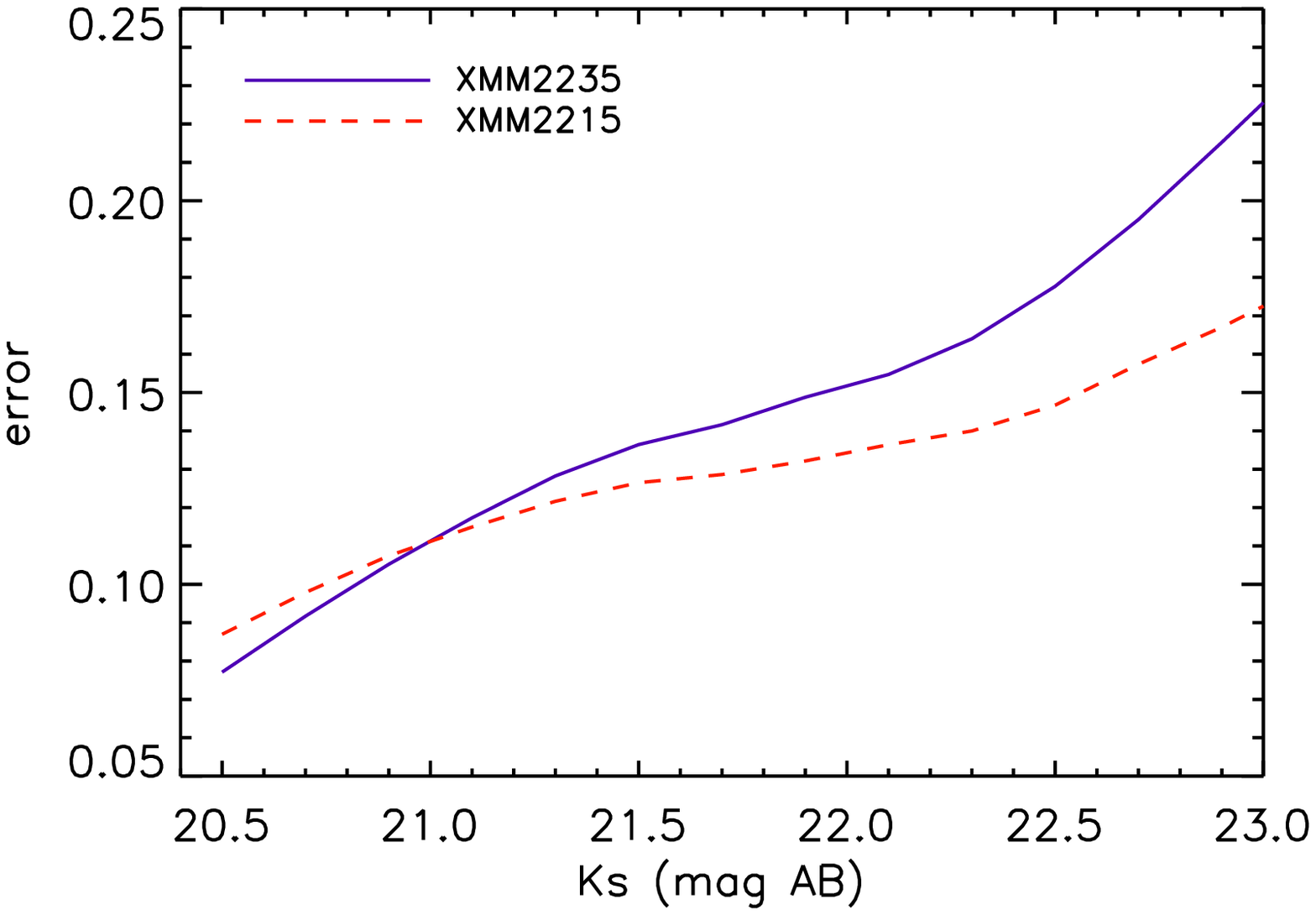}} \\
\end{tabular}
\caption{Photometric errors corresponding to the scatter between input and output magnitudes in simulations for each cluster in the bands used for colours. In this way, we estimate a maximal error on aperture magnitude in each bandpass.}
\label{errphot}
\end{figure*}

Previous works have shown that photometric errors estimated by Sextractor are underestimated \cite[e.g., ][]{benitez04, giavalisco04,raichoor11}. We therefore estimate photometric errors on the aperture magnitudes through simulations. For each filter, we simulate $10,000$ galaxies between $20$ and $25$~mag and then drop the simulated objects in a similar background than the one of the real image (see section~\ref{sec:sizes} for more details on the simulations). Simulated galaxies are then recovered with Sextractor. We compare input and output magnitudes and we estimate the photometric errors as the scatter in magnitude bin of 0.2~mags. For the ACS filters, typical errors on colours within the effective radius are around $\sim 0.03-0.07$~mag whereas Sextractor errors are in average $\sim 0.02$~mag. All errors are summarized in figure~\ref{errphot}. As already shown in previous work \cite[e.g., ][]{benitez04, giavalisco04,mei09}, we also found a systematic shift of 0.2~mag between input and recovered Sextractor \texttt{MAG\_AUTO} magnitudes. 

\subsection{Sizes}
\label{sec:sizes}

We use GALAPAGOS \citep{barden05} to estimate the sizes of all detected galaxies in the ACS/HST $i_{775}$-band for the two closest clusters RX0152 at $z=0.84$ and RCS2319 at $z=0.91$ and the $z_{850}$-band for the others.  Our sizes are therefore derived in the B rest-frame band for all clusters. We notice that we do not expect significant differences ($\sim20\%$) with sizes estimated in redder bands as demonstrated by \cite{cassata11, damjanov11,newman10, szomoru13}. 

GALAPAGOS is an IDL based pipeline to run Sextractor and GALFIT (v3.0.2, \citealp{peng02}) specially designed to be used on large datasets. On each detected source, GALFIT models a galaxy light profile using a 2D Sersic profile \citep{sersic68} with a fixed sky value previously estimated by GALAPAGOS. We let the default constraints on the Sersic index $n$, the effective radius $r_e$, the axis ratio $q$, the position angle P.A. and the magnitude to run GALFIT: $ 0.2 < n < 8$, $ 0.3 < r_e < 750$~pix, $ 0.0001 < b/a < 1$, $ -180^\circ < P.A. < 180^\circ$, $ 0 < mag < 40$ and $ -5 < \delta mag < 5$ and use a synthetic PSF from Tiny Tim \citep{krist11}.

 The sky is fixed and measured by GALAPAGOS before running GALFIT in a $3 \times$ enlarged isophotal stamp. GALAPAGOS uses a flux growth method to estimate the sky around an object. It calculates the average flux in an elliptical annuli centered on the object excluding other detected sources to obtain the flux as a function of radius. Once the slope levels off, it determines the sky from the last few annuli. 
 

In the following, we use as primary size estimator the circularized effective radius defined by:
\begin{equation}
R_{\rm eff} = r_e \times \sqrt{b/a}.
\end{equation}

\begin{figure*}
\begin{center}
\includegraphics[width = 0.80\textwidth]{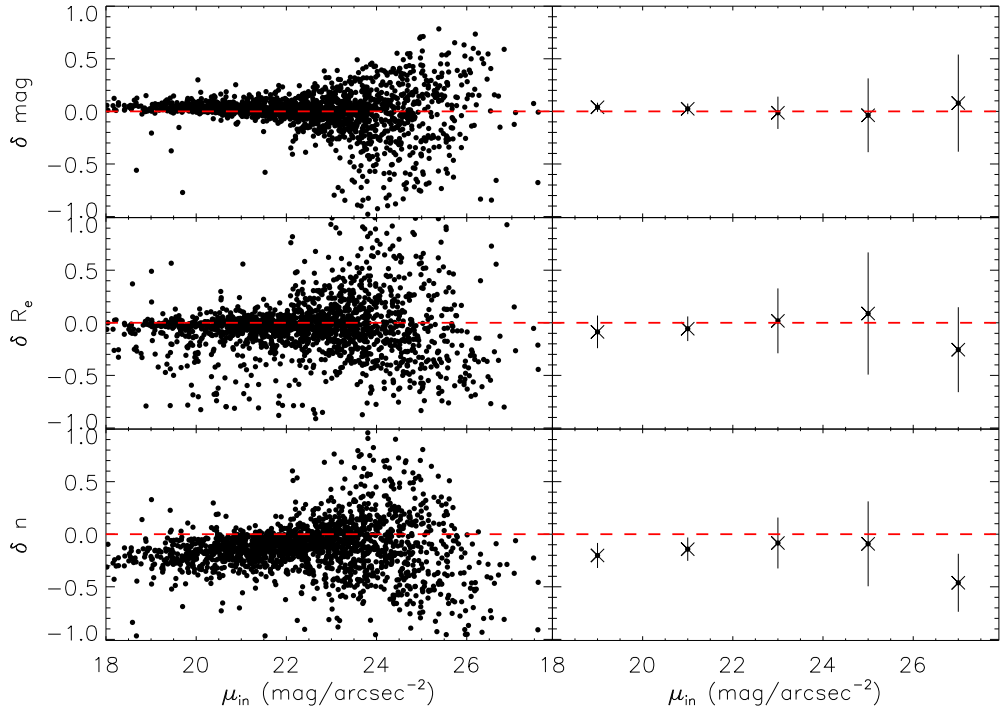}
\caption{In the left pannels, difference between the estimated parameter by GALFIT and the input parameter in function of the surface brightness (as follows $\delta r_e = (r_{e,out} - r_{e,in})/r_{e,in}$, $\delta mag = (mag_{out} - mag_{in})$, $\delta n = (n_{out} - n_{in})/n_{in}$), and in the right pannels, mean value and scatter estimated in different bins with the $3\sigma$-clipping method.}
\label{testgala}
\end{center}
\end{figure*}

The accuracy of our size estimates is assessed through extensive simulations in which we drop mock galaxies in real background images. The background is built as a composite image of empty regions distributed in all the fields. We generate $3000$ galaxies with random magnitudes in the range $20 < z_{850} < 26$~mag, and a Sersic profile with random effective radii, Sersic indices and ellipticities. These properties are taken randomly following the real distributions: effective radius distribution peaks at $\langle r_e \rangle = 0.4$~arcsec with a dispersion of $\sigma_{r_e} = 0.24$, Sersic indices peaks at $\langle n \rangle = 3.7$ with $\sigma_n = 1.5$, $\langle e \rangle = 0.67$ with $\sigma_e = 0.17$, and magnitudes peaks at $z_{850} = 24$~mag with $\langle z_{850} \rangle = 1.5$. A Poisson noise is added and the simulated galaxy is convolved with a PSF.  We then run GALAPAGOS on the mock dataset and compare the output and input parameters. Results are shown in figure~\ref{testgala} and table~\ref{tabletest} as a function of the input surface brightness, $\mu_{in} = mag_{in} + 2.5 \log(2 \pi r_{e,in}^2)$~mag.arcsec$^{-2}$ for the $z_{850}$ band images. 
Our main conclusion after inspection of figure~\ref{testgala} is that results are robust for objects brighter than $24$~mag/arcsec$^2$. Sizes can be recovered with a systematic error lower than $10\%$ and a dispersion lower than $30\%$ up to $\mu = 24$~mag/arcsec$^2$ (see Table~\ref{tabletest} for errors details). Similar conclusions hold when the $i_{775}$ band is used instead of the $z_{850}$ band. 

Since the size measurements are very sensitive to the sky estimate, especially in dense regions such as clusters we double checked the robustness of our size estimates by running GALFIT a second time with a sky value estimated with the method described in  \cite{raichoor12}. We find that both methods deliver consistent size measurements at $1\sigma$ level. 

\begin{table*}
\begin{center}
  \begin{tabular}{ c  c  c  c  c  c  c  c }
    \hline
    \hline
bin $\mu_{in}$ &  $n_{obj}-3\sigma$ & $\langle \delta r_e \rangle$ & $\sigma(\delta r_e)$ & $ \langle \delta mag \rangle$ & $\sigma(\delta mag)$ & $\langle \delta n \rangle$ & $\sigma(\delta n)$  \\
\hline
$[18, 20]$   &      175 & -0.08  &  0.16   &   0.04 &  0.04 &  -0.19  &  0.12  \\
$[20, 22]$   &      674 & -0.05  &  0.12   &   0.02 &  0.05 &  -0.14  &  0.12  \\
$[22, 24]$   &     1307 &  0.02  &  0.30   &  -0.01 &  0.17 &  -0.09  &  0.28  \\
$[24, 26]$   &      903 &  0.12  &  0.64   &  -0.04 &  0.36 &  -0.08  &  0.43  \\
$[26, 28]$   &       78 & -0.21  &  0.44   &   0.12 &  0.40 &  -0.35  &  0.36  \\
    \hline
  \end{tabular}
\caption{Bias and dispersions in the results of simulated quantities ($r_e$, $mag$, $n$) for {\bf different surface brigthness bins (left column).}}
\label{tabletest}
\end{center}
\end{table*}

\subsection{Stellar masses}

We estimate stellar masses of our galaxies through SED fitting using the spectral library of \cite{BC03} (hereafter, BC03) with the LePhare code \citep{arnouts99, ilbert06}. We consider galaxy templates from stellar population models with a \cite{chabrier03} IMF, 3 different metallicities ($Z = 0.004$, $Z=0.008$ or $Z=0.02$), exponentially declining star formation histories (SFH) $\psi(t) \propto e^{-t/\tau}$ with a characteristic time $0.1 \leq \tau \; \text{(Gyr)} \leq 30$, and no dust extinction. The redshift of the galaxy is fixed to the cluster redshift before performing the fit, to avoid degeneracies between redshift and stellar mass. We use \texttt{MAG\_AUTO} magnitudes in all available filters ($i$, $z$, J and K for RX0152, $r$, $i$, $z$ and K) with errors estimated as explained in section~2.2.



\subsection{Morphologies}
\label{sec:morpho}
Deriving morphologies of $z>1$ galaxies remains a challenge even with the high spatial resolution delivered by the HST.  Therefore, in this work, we estimate B-rest frame morphologies visually and with an automated method. We only derive morphologies for galaxies with $z_{850}<24$~mag since a visual inspection and also preliminary tests with our automated algorithms indicate that galaxies fainter than this magnitude have a signal-to-noise ratio too low to derive reliable classifications (see also, e.g. Postman et al. 2005). 

\subsubsection{Automated morphologies}  

For the automated morphological classification, we use GalSVM, a non-parametric code based on support-vector machines \citep{huertas08, huertas09, huertas11}. The code follows a bayesian approach to associate a probability to each galaxy to be of a given morphological type, previously defined by the user. GalSVM is trained on a local sample  with known visual morphologies chosen at the same rest-frame band than the high redshift sample. The training set is then moved at high redshift (which includes image degradation, resampling...) and dropped in the high-z real background. The code measures afterwards a set of morphological parameters (asymmetry, concentration, smoothness, ...) on the simulated dataset and trains a support vector machine. During the classification process, possible systematic errors detected in the testing step are taken into account. The local sample used in this work is a catalogue from the Sloan Digital Sky Survey DR7 of about 14,000 galaxies visually classified \citep{nair10}.  For each cluster, we took a sample of 3500 galaxies and used 3000 for training and 500 to estimate errors.  

\begin{figure}
\includegraphics[width = 0.49\textwidth]{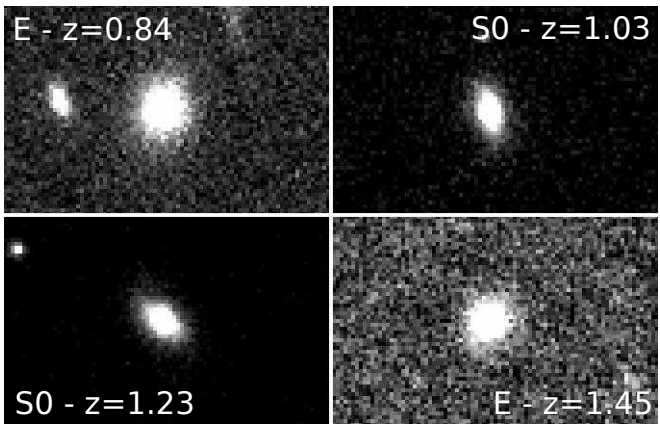}
\caption{Early-type galaxies in four different clusters.}
\label{morphoE}
\end{figure}

In this work, galaxies are classified into three morphological classes (ellipticals, lenticulars and spirals/irregulars). We refer the reader to Huertas-Company et al. (2013) for more details on how these classes are defined. However, in the following we will only focus on ETGs (including ellipticals and lenticulars) defined as those objects with  $P(ETG)>0.5$ (some example stamps are shown in figure~\ref{morphoE}).  


\subsubsection{Visual morphologies} 

Three of us (LD, MHC and SM) also visually classified all the sample in the same three morphological classes (elliptical, lenticular and spiral/irregular). For the final visual classification we only keep objects for which at least two classifiers agree. A general good agreement is found between visual and automated classification of ETGs, i.e. we measure between 2 and 15\% disagreement (depending on the cluster - see table~\ref{Nsample}) which corresponds to an average of $5\pm2\%$ discrepancy in the whole cluster ETGs selection. 

The level of discrepancy between ellipticals and lenticulars visually and automatically classified is logically higher and reaches $\sim30\%$. Interestingly, this is roughly the same level of agreement expected between two independent human classifiers \citep[see also][]{postman05}.

We also compared our morphologies to published results. Two of the clusters (XMM1229 and XMM2215) have indeed available visual morphologies for some objects which we have compared to our automated determination:

\begin{itemize}
\item \cite{santos09} visually classified  26 galaxies in XMM1229. Our automated classification agrees at an $85\%$ level with their results. Only 4 galaxies have an associated probability smaller than 0.5 and are visually classified as ETGs by \cite{santos09}.
\item A similar study was done in XMM2215, the most distant cluster in our sample at $z=1.45$, by \cite{hilton09} who visually classified 36 galaxies with $z_{850} < 24$~mag. For that particular dataset, we find that $22\%$ ($8$ galaxies) of the objects have different classifications. A similar level of disagreement ($14\%$ ($5$)) is however measured between our visual classification and the published one.
\end{itemize}

We will discuss how the differences between the different morphological classifications affect our main results in section~\ref{sec:size_evol}.


\subsection{Sample selection}
\label{sec:selection}
 In the remaining of this work we will consider only galaxies with $z_{850}<24$~mag in order to have accurate size estimates and morphologies (see sections~\ref{sec:sizes} and~\ref{sec:morpho}). Several further selections to build our final sample of cluster members are done, which are detailed in the following. 

\begin{figure*}
\begin{center}
  \begin{tabular}{c}
	\subfloat[]{\includegraphics[width = 0.35\textwidth, trim = 0.0cm 0.4cm 0.2cm 0.2cm, clip]{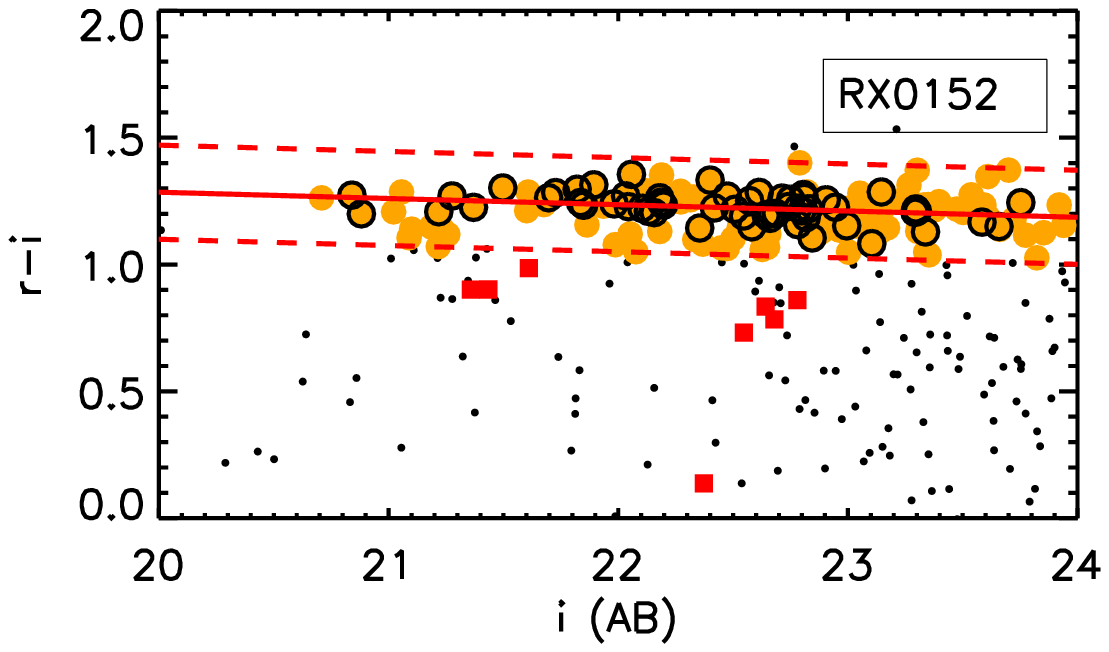}} \\
	\subfloat[]{\includegraphics[width = 0.72\textwidth, trim = 0.0cm 0.1cm 0.05cm 0.0cm, clip]{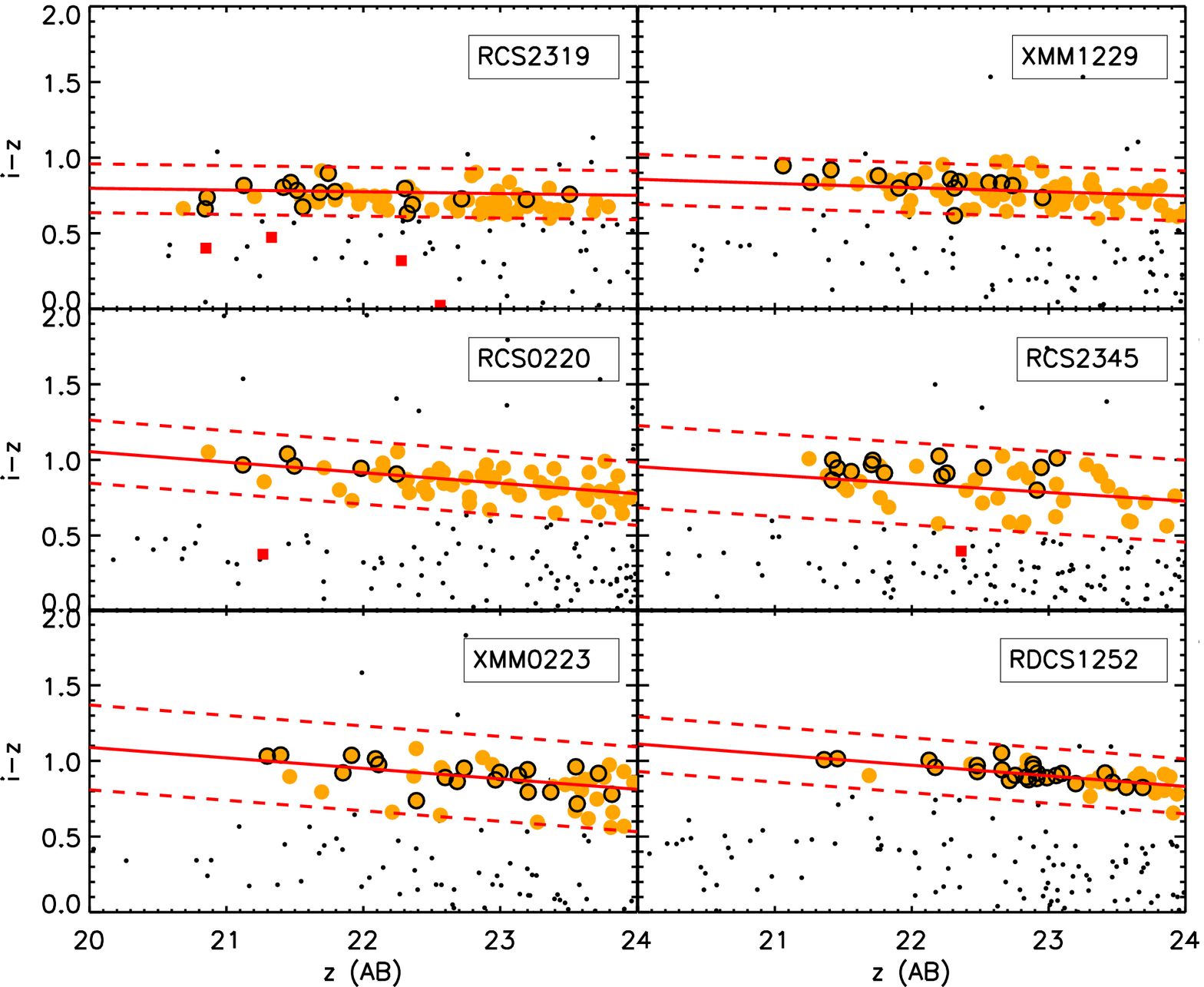}} \\
	\subfloat[]{\includegraphics[width = 0.70\textwidth, trim = 0.0cm 0.3cm 0.05cm 0.1cm, clip]{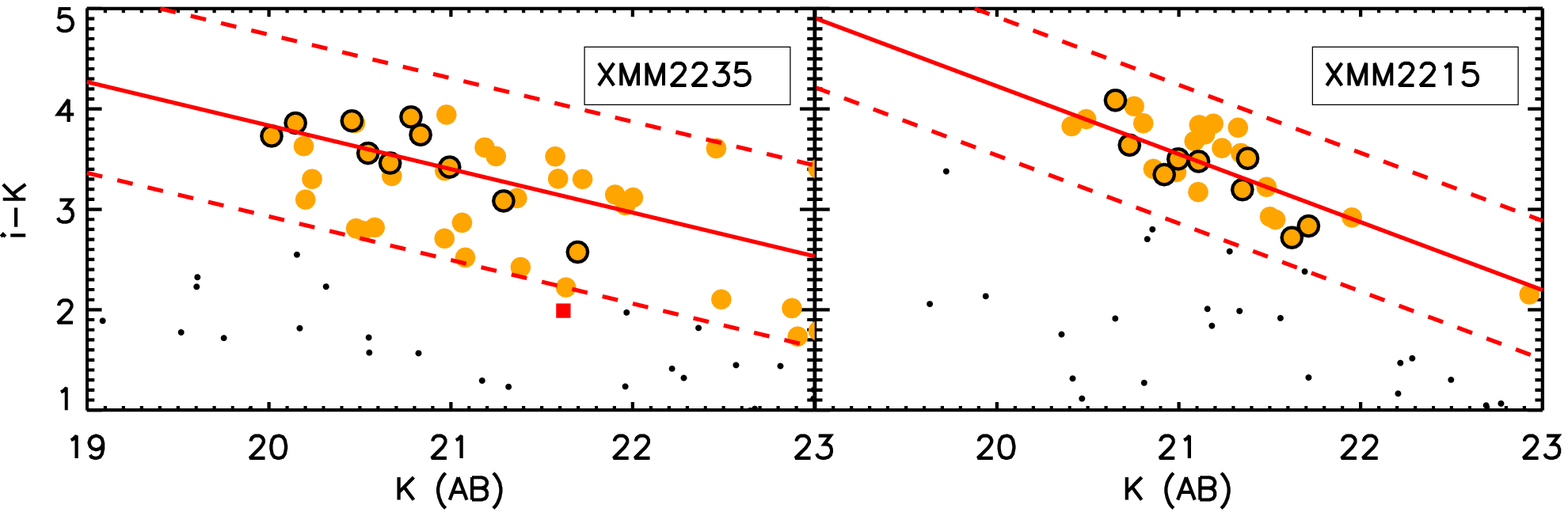}} \\
  \end{tabular}
\caption{Color-magnitude diagram: $(r_{625} - i_{775})$ vs $i_{775}$ for RX0152 galaxies, $(i_{775}-z_{850})$ vs $z_{850}$ for cluster galaxies between $z=0.9$ and $z=1.23$, $(i_{775}-K_s)$ vs $K_s$ for the highest cluster galaxies at $z=1.39$ and $z=1.45$. Orange circles with black contours correspond to ETGs with spectroscopic redshift in the cluster used to fit the red sequence (red line). Red dashed lines correspond to the fitted red sequence at $\pm 3 \sigma$. Orange circles correspond to the selected ETGs on the red sequence $\pm 3\sigma$. Red squares are for ETGs with spectroscopic redshift in the cluster, but not on the red sequence.}
\label{redseq}
\end{center}
\end{figure*}

\begin{figure}
\includegraphics[width = 0.49\textwidth]{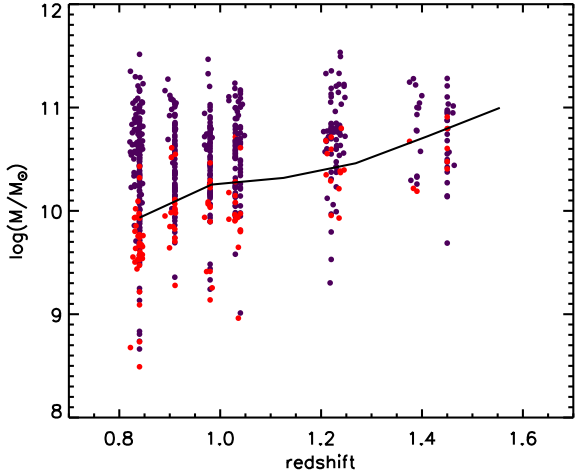}
\caption{Stellar mass as a function of redshift of the HCS ETG sample in the z-band images. Red dots are $M^{lim}$ and the black line shows the 80\% completeness level \citep{pozzetti10}.}
\label{compl_mass}
\end{figure}

\begin{figure*}
	\subfloat[]{\includegraphics[width = 0.49\textwidth]{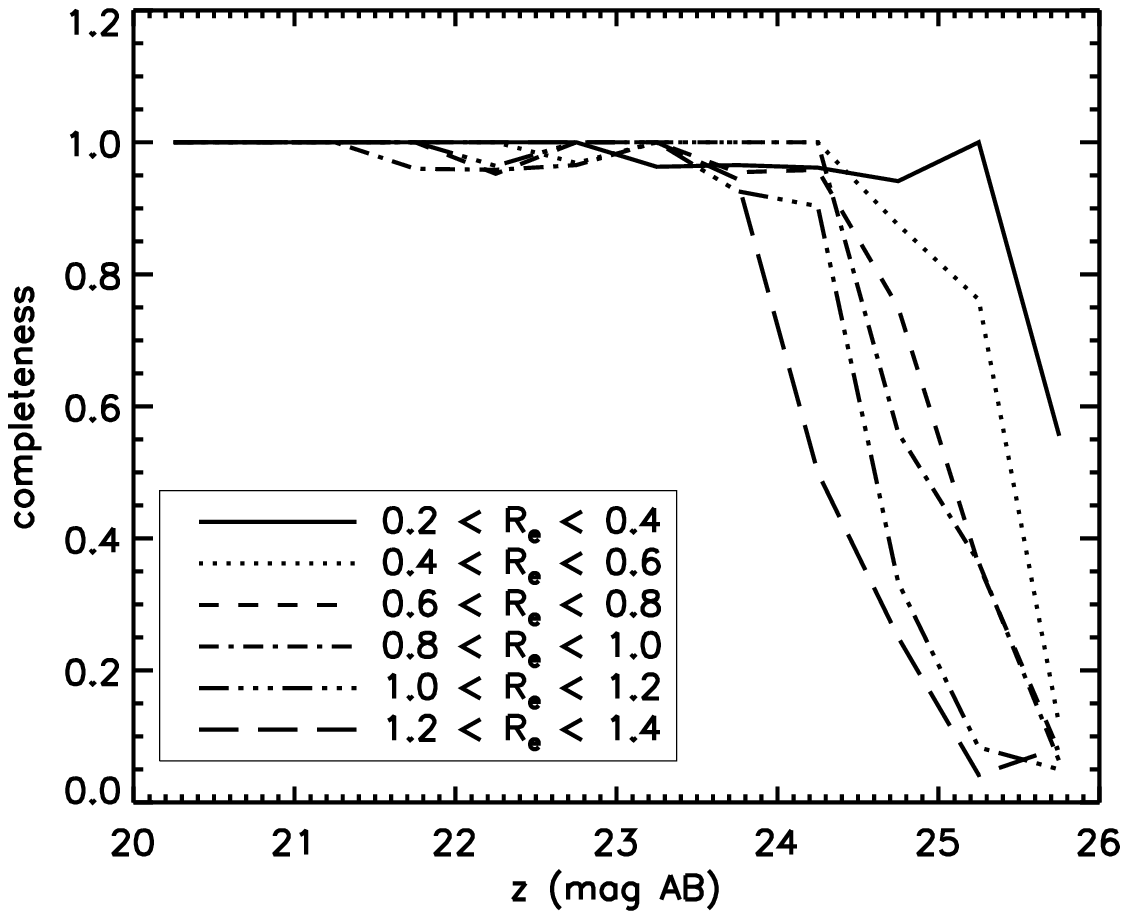}}
	\subfloat{\includegraphics[width = 0.49\textwidth]{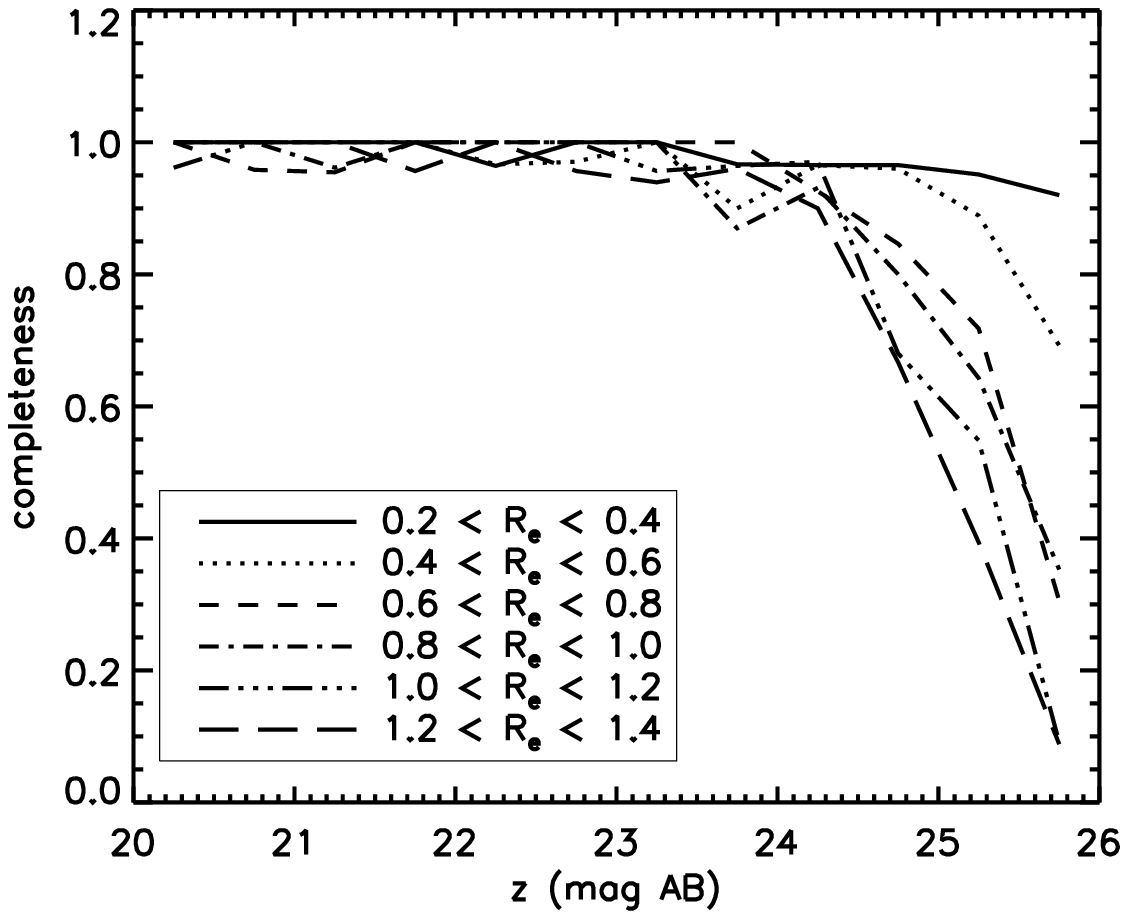}}
	\caption{Completeness of the HCS sample in the z-band images as a function of magnitude and size (in arcsec) : \textit{left pannel}: for the least deep image (RCS2319) and \textit{right pannel}: for a typical image depth of HCS (here, XMM1229).}
\label{compl_re}
\end{figure*}

\begin{enumerate}
	\item Since we are interested in passive ETGs, we use the red-sequence to determine cluster members when no spectroscopic redshift is available. We therefore selected objects belonging to the cluster according to their position in the observed color-mag plane closer to the rest frame $(U-B)$ versus $B$ plane. The magnitudes used change therefore from cluster to cluster depending on the redshift (fig.~\ref{redseq}). Colors are measured as explained in section \ref{sec:photo}, i.e. within an aperture of one effective radius. For each cluster, we then fit a linear red sequence ($color = a + b \times mag$), using only spectroscopically confirmed members and then select cluster members within $3\sigma$ of the best fit (fig.~\ref{redseq}). The fit is performed with an iterative sigma-clipping linear regression and the scatter $\sigma$ is computed with a robust standard deviation based on bi-square weights \citep[Tukey's biweight][]{press92}.  We measure a fraction of outliers, corresponding to the fraction of galaxies with a spectroscopic redshift outside the cluster, between $5$ and $20\%$ (see details in table~\ref{Nsample}), which are removed from the final selection. Only two clusters, RCS2319 and XMM2235, have larger contaminations ($31\%$ and $44\%$ respectively). 

		An alternative to the red-sequence based selection is a selection based on photometric redshifts, which has in principle the advantage of selecting all members independently of their star formation activity. We therefore obtained photometric redshifts for all the detected sources through SED fitting with two different codes: LePhare \citep{arnouts99, ilbert06} and EAZY \citep{brammer08}. For LePhare, we used synthetic galaxy templates from Bruzual and Charlot (2003) models with a \cite{chabrier03} IMF, three different metallicities ($Z = 0.004$, $Z=0.008$ or $Z=0.02$), exponentially declining star formation histories (SFH) $\psi(t) \propto e^{-t/\tau}$ with a characteristic time $0.1 \leq \tau \; \text{(Gyr)} \leq 30$, and no dust extinction \citep[e.g., ][]{ilbert06}. For EAZY, we kept default settings and a K-band magnitude prior. We then consider that a galaxy belongs to a given cluster if $\left | z_{phot}-z_{cluster}\right | <\Delta z$ with $\Delta z$ changing from cluster to cluster to maximize the completeness and minimize the contamination simultaneously based on the spectroscopic sample only as described in \cite{2009A&A...508.1173P}.  When no color preselection is made, the average level of contamination is very high, $\sim40\%$, which is probably due to the fact that our sample lacks of blue filters. If instead, we restrict to red galaxies by applying a colour cut ($r_{625}-i_{775}>0.8$, $i_{775}-z_{850}>0.5$ and $i_{775}-K_s>1.8$), we still find large contaminations ($\sim 40-50\%$) with the 2 algorithms for the two most distant clusters (XMM2235 and XMM2215) and a contamination around $~15-35\%$ for $z<1.3$ clusters. These values are still larger than what is obtained with the red-sequence selection, so we decided not to use photometric redshifts for selecting cluster members in this work. 

\item Among the selected red-sequence population we then select ETGs based on our automated classifications as described in section~\ref{sec:morpho}. {\bf A selection based on visual morphologies leads to similar results given the }

\item We also remove objects for which the Sersic fits did not converge (see section~\ref{sec:sizes} for size determination method). We consider that the fitting procedure has converged if $mag \leq 24$, $|M_{galfit} - M_{SEx}| < 0.8$, $0.1 < R_{eff} < 1.6"$ and $n \ne 8$. We obtain less than $1\%$ non converged fits for cluster ETGs selected on the red sequence with $\log(M/\text{M}_{\odot}) \geq 10.5$ and $z_{850} < 24$~mag. Exact numbers are detailed in table~\ref{Nsample}. This number is negligible compared to the total number of selected passive ETGs, so it has no impact in our results.

\item Finally, we keep only ETGs with a stellar mass greater than $3\times10^{10} M_{\odot}$  to keep a complete sample. We used two different approaches to estimate the mass completeness. First, following \cite{pozzetti10}, for each passive ETG with spectroscopic redshift, we compute the limiting stellar mass ($M^{lim}$) given by $\log(M^{lim}) = \log(M) - 0.4 (z_{850} - z_{lim})$  where $z_{lim} = 24$~mag in our case. We use the limiting mass of the $20\%$ faintest galaxies at each cluster redshift and estimate that way the mass limit at $80\%$ completeness. As shown in figure~\ref{compl_mass}, our sample is 80\% complete for galaxies with stellar mass greater than $\log(M/\text{M}_{\odot}) = 10.2$ at $z \sim 1$ and $\log(M/\text{M}_{\odot}) = 10.8$ at $z \sim 1.45$. Second, we used an approach similar to \cite{bundy10} and estimated the apparent magnitude in the z-band of a typical passive galaxy using stellar population models (i.e. solar metallicity and no dust and with a $\tau = 0.5$~Gyr burst of star formation occurring at $z_f = 5$). At $z=1.45$, the redshift of the most distant cluster, a galaxy of $z_{850}=24$~mag has a stellar mass of $log(M/\text{M}_{\odot}) = 10.7$ which is roughly consistent with the estimate. We also make sure that, with the depth of our images, we detect $>90\%$ of the galaxies  with $z_{850} < 24$~mag through simulations (fig.~\ref{compl_re} independently of their size).

\end{enumerate}

The final cluster sample contains $319$ cluster galaxies, among which $149$ are spectroscopically confirmed members. Details are given in table~\ref{Nsample}.

\begin{table*}
\begin{center}
\begin{tabular}{ c c c c c c c }
\hline
\hline
Cluster  & \#ETG without z & \#ETG with z & contamination &  \#bad fit& \#LTGs & \#E-S0 \\
\hline
 RXJ0152   &  50 & 47 &   14\% (7)  &  4 &  4-6\%  (4+2)  & 30\%  \\
 RCS2319   &  20 & 13 &   31\% (6)  &  1 & 6-12\%  (2+2)  & 36\% \\
 XMMJ1229  &  24 & 15 &   12\% (2)  &  0 &  3-8\%  (1+2)  & 21\% \\
 RCS0220   &  28 &  5 &    -- (0)   &  0 & 9-15\%  (3+2)  & 31\% \\
 RCS2345   &  12 & 12 &   20\% (3)  &  1 & 0-4,5\% (0+1)  & 23\% \\
 XMMJ0223  &  12 & 19 &    5\% (1)  &  0 &   10\%  (3+0)  & 32\% \\
 RDCSJ1252 &   4 & 25 &   14\% (4)  &  0 &    4\%  (1+0)  & 31\% \\
 XMMU2235  &   4 &  9 &   44\% (7)  &  1 &  0-8\%  (0+1)  & 15\% \\
 XMMJ2215  &  11 &  8 &   20\% (2)  &  0 & 5-15\%  (1+2)  & 20\% \\
\hline
\end{tabular}
\caption{Number of galaxies in the final sample of the nine clusters with $\log(M/\text{M}_{\odot})>10.5$ and $z_{850} < 24$. \#ETGs without z: total number of early-type galaxies without spectroscopic redshift on the red sequence (RS) with $P(ETG) > 0.5$, \#ETG with z: total number of early-type galaxies with spectroscopic redshift on the RS with $P(ETG) > 0.5$, contamination : percentage of spectroscopic confirmed outliers from cluster among ETGs on the RS (number of galaxies), \#bad fit: number of ETGs for which Galfit does not converge, \#LTGs: percentage of misclassified late-type galaxies by GalSVM (number of clear LTGs + number of uncertain LTGs), \#E-S0: percentage of disagreement between visual and automated GalSVM classification of E and S0 galaxies.}
\label{Nsample}
\end{center}
\end{table*}

\section{Field comparison sample}
\label{sec:field}

In order to disentangle the environmental effects on the size evolution of passive ETGs, we define a field sample from a combination of four different datasets to be compared with our main cluster sample.

\begin{enumerate}

\item A first set of galaxies is built by putting together all foreground and background galaxies detected in the clusters fields with spectroscopic redshifts ($|z - z_{cl}| > 0.02$) in the redshift range $0.7 < z < 1.6$. We then apply the same colour selection than for cluster galaxies (see section~\ref{sec:selection}). This ensures a subsample with exactly the same properties in terms of resolution and depth than the main cluster sample. All derived quantities (stellar masses, sizes and morphologies) are therefore obtained with the same methods described for the cluster sample in section~\ref{sec:clusters}. This first field sample is referred in the following as the \emph{HCS field sample} and contains 30~galaxies.

\item To increase the number of field galaxies, we add a sample of galaxies from the COSMOS survey \citep{scoville07} (referred in the following as \emph{the COSMOS sample}) with photometric redshifts between $z=0.7$ and $z=1.6$ from \cite{george11} and sizes from \cite{huertas12}. Passive galaxies are selected using the colour selection $\texttt{NUV}-R > 3.5$ \citep{ilbert10} corrected from dust extinction where \texttt{NUV} is the near ultraviolet band from Galex and $R$ is an optical band from Subaru Telescope. Sizes are estimated using also GALAPAGOS on the HST/ACS F814W (i-band) images as described in \cite{huertas12}. Since galaxies in the COSMOS sample are selected based on the i-band magnitude ($i<24$), we have checked the reliability of the size estimates in the redshift range explored by carrying out similar simulations than for the main sample. We find comparable results than for the HCS objects brighter than $z_{850}=24$~mag \citep[i.e. systematic error lower than $0.1$ and a reasonable scatter lower than $0.2$ up to $i < 24$~mag][]{huertas12}. However, the i-band selection implies a mass completeness close to $10^{11} \; \text{M}_{\odot}$ at $z \sim 1.5$ \citep[extrapolation of figure 4 of][]{huertas12}, and the COSMOS sample is therefore less complete than our main cluster sample (see section~\ref{sec:selection}). We will discuss the effects of this when discussing our main results. Stellar masses are estimated using the LePhare software with BC03 library and a Chabrier IMF with all the available filters in COSMOS with the same parameters used for the HCS cluster and field samples. Our stellar mass estimates show in fact a small shift (0.2~dex) when compared to the \cite{bundy06} stellar mass estimates used in \cite{george11}. These small shifts are common when comparing stellar masses used with different algorithms and settings. Finally morphologies were derived automatically \citep[see][]{huertas12} and visually by two of us (LD and MHC) following the same methodology than for the main sample. The whole COSMOS sample contains 211 galaxies.

\item Additional field galaxies in the redshift range $1.1<z<1.4$ with published sizes, stellar masses and morphologies \citep{raichoor12} from the GOODS-CDF-S field \citep{giavalisco04} are also considered (\emph{GOODS sample} in the following).The sample is selected from the public GOODS-MUSIC v2 catalog \citep{santini09} and at $z_{850}=24$~mag, the sample is more than $70\%$ complete. We select only red galaxies with $0.75 < i-z < 1.1$. All these galaxies have a spectroscopic redshift. We refer to \cite{raichoor11} for a detailed description of the selection and for further informations about the completeness of the sample. Stellar masses in the GOODS-S sample were measured with an SED fitting code (different from LePhare) using BC03 stellar population models and a Salpeter IMF \citep[see][for details]{raichoor11}. To convert into a Chabrier IMF, we applied the following correction: $\log(M_{\text{Chabrier}}) = \log(M_{\text{Salpeter}}) - 0.25$ taken from \cite{bernardi10}. We have checked that the resulting stellar masses are consistent at $1\sigma$ level with the ones obtained with LePhare. Sizes were computed on the HST/ACS $z_{850}$ image using GALFIT with a fixed sky value previously derived on a larger stamp centered on the ETG (see section 3 of \citealp{raichoor12} for for more details on the method). Finally galaxies were visually classified (E/S0 types) in the HST/ACS F850LP images as described in \cite{mei12}. This sample contains 17 galaxies.

\item Finally, the field sample also contains galaxies in the redshift range $0.7<z<1.6$ from the CANDELS survey with published redshifts and sizes by \cite{newman12} (referred in the following as \emph{the CANDELS sample}). Sizes were also derived with GALFIT (see \cite{newman12} for details) in the optical rest-frame band. We recomptuted stellar masses with LePhare through SED fitting using \cite{BC03} models and a Chabrier IMF in the same way as the cluster sample instead of using the published ones. Galaxies in this sample were selected to be quiescent (SSFR$<0.02 \; \text{Gyr}^{-1}$ and no MIPS detection), and have been visually classified by us. The CANDELS sample is complete for stellar masses $log(M/\text{M}_{\odot}) > 10.52$ with a Chabrier IMF and BC03 model (see section 2.4 of \cite{newman12} for details). We have 125 galaxies in this sample.
\end{enumerate}

Our final field sample contains $383$ galaxies. Details are given in table~\ref{Fsample}.

\begin{table*}
\begin{center}
\begin{tabular}{ c c c c c }
\hline
\hline
redshift bin &  \# HCS &  \#  COSMOS &  \# GOODS-S  &   \# CANDELS  \\
\hline
$[0.7, 0.9]$ &   9 & 91  & ... & 23 \\
$[0.9, 1.1]$ &  12 & 83  & ... & 40 \\
$[1.1, 1.6]$ &   9 & 37  &  17 & 62 \\
\hline
\end{tabular}
\caption{Number of field galaxies in the final sample with HCS and COSMOS data, GOODS sample from Raichoor et al. (2012), and CANDELS sample from Newman et al. (2012) } 
\label{Fsample}
\end{center}
\end{table*}

\section{Results}

\subsection{Super dense galaxies at $z\sim1$}

We study first the fraction of \emph{compact} objects  in the two different environments without morphological selection. \cite{poggianti12} measured a clear difference in the fraction of the so-called \emph{super dense galaxies (SDGs)} in clusters ($\sim20\%$) and in the field ($4\%$) in the local universe without morphological distinction. We compare our results at high redshift, by taking the same selection for SDGs in the same stellar mass range: 
\begin{equation}
\Sigma_{50} > 3\times 10^9 \; \text{M}_\odot \text{kpc}^{-2}
\end{equation}
where the mean mass surface density is defined by $\Sigma_{50} = 0.5 M_* / \pi R_e^2$, in the stellar mass range $10.5 < \log(M/\text{M}_\odot < 11.6$ with no morphological selection (all figures in this work show only ETGs).  
 For this comparison, we restrict the cluster sample to spectroscopically confirmed members ($212$~galaxies) to include all morphological types in the selection (not only ETGs), and the field sample is limited to the HCS sample, with $122$~galaxies, because it is the only one for which the sample is not selected based on morphology.
We find then 67~SDGs live in clusters ($32^{+4}_{-3}~\%$) and $26$~SDGs in the field ($21^{+4}_{-3}~\%$).

The fraction of SDGs in clusters is only  $1.5$ larger than in the field at $z\sim1$, which is $3$ times less than the difference found by \cite{poggianti12} in the local Universe. Concerning the morphological properties of SDGs we find that the fractions of late-type SDGs is $\sim2$ times higher in our high redshift sample and comparable to the values measured by \cite{valentinuzzi10b} at similar redshifts in the ESO Distant Cluster Survey \citep{white05}. Effective radius, Sersic indices, axis ratios and stellar masses of our cluster and field SDGs are consistent within $1\sigma$ with the local values of \cite{poggianti12}. Table~\ref{SDG_prop} summarizes all the properties of the SDGs found in clusters and in the field compared to the ones reported by \cite{poggianti12} in the local universe.

\begin{table*}
\begin{center}
\begin{tabular}{c c c c c }
\hline
\hline
 & \multicolumn{2}{c}{at $z \sim 1$} & \multicolumn{2}{c}{at $z\sim0$} \\
\hline
& Cluster (HCS)  & Field (HCS)  &   Cluster (WINGS)   &  Field (PM2GC)\\
\hline
$f_{SDGs}$ & $32^{+4}_{-3}\%$ & $21^{+4}_{-3}\%$ & $17~\%$  &  $4.4~\%$\\
$\langle R_e \rangle$ & $1.55 \pm 0.08$ & $1.38 \pm 0.08$ & $1.57 \pm 0.34$  &  $1.45 \pm 0.26$  \\
  $\langle n \rangle$ & $2.98 \pm 0.13$ & $2.61 \pm 0.26$ & $3.1 \pm 0.8$ & $2.8 \pm 0.6$ \\
$\langle b/a \rangle$ & $0.61 \pm 0.04$ & $0.52 \pm 0.05$ & $0.65 \pm 0.16$ & $0.48 \pm 0.13$ \\
$\langle \log (M/\text{M}_\odot) \rangle$ & $10.87 \pm 0.06$ & $10.90 \pm 0.05$ & $10.96 \pm 4.33$ & $10.78 \pm 3.41$ \\
				$f_{ELL}$ & $24 \pm 6~\%$ & $15^{+11}_{-7}~\%$ & $29.1 \pm 7.8~\%$ & $22.7 \pm 7.2~\%$ \\
				$f_{S0}$ & $50 \pm 7~\%$ & $65^{10}_{12}~\%$ & $62.0 \pm 10.7~\%$ & $70.5 \pm 12.7~\%$\\
				$f_{LTG}$ & $19^{+6}_{-5}~\%$ & $19^{10}_{8}~\%$ & $8.8 \pm 4.4~\%$ & $6.8 \pm 3.9~\%$ \\
 $f_{unknown}$ & $4^{+4}_{-2}~\%$ & ... & ... & ... \\
\hline
\end{tabular}
\caption{SDG properties in clusters and in the field. For high redshift SDGs, mean values are computed with $3\sigma$-clipping method and errors by bootstrapping. Values for local SGDs are taken from \protect\cite{poggianti12}.}
\label{SDG_prop}
\end{center}
\end{table*}

\subsection{The mass-size relation of early-type galaxies in clusters at $z\sim1$}
\label{sec:results}

\begin{figure*}
\includegraphics[width = 0.98\textwidth]{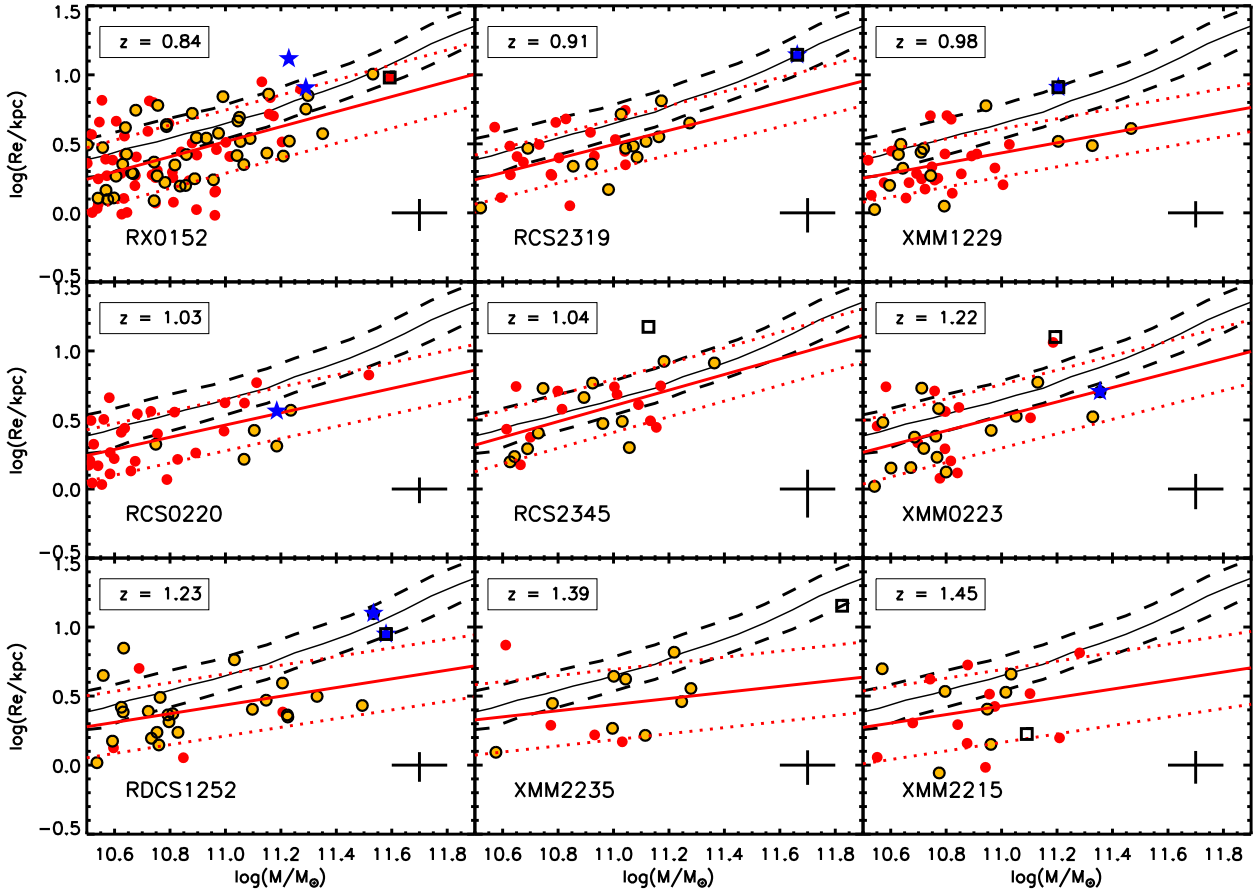}
\caption{Mass-size relation of passive ETGs in clusters. Orange circles with black contour represent galaxies with spectroscopic redshift, red circles are the red sequence sample and the blue stars correspond to the central dominant galaxies we manage to identify whereas the black squares correspond to the BCGs identified by Lidman et al. (in prep.). The black solid line corresponds to the local relation of \protect\cite{bernardi12} and the $1\sigma$ standard deviation in black dashed lines. Each red line corresponds to the fit for each cluster sample with the $1\sigma$ standard deviation in red dotted line. In the right end corner of each pannel, the black cross represents the median error bar on mass and size.}
\label{MSR}
\end{figure*}

From this point we focus only on the passive ETG population. In figure~\ref{MSR}, we show the mass-size relation (MSR) of passive ETGs of the nine studied clusters separately. The figure also shows the best-fit power law model  $\log(R_{\text{eff}}/kpc) = \kappa + \beta \times \log(M/\text{M}_{\odot})$ for each cluster with  $ 10.5 <log(M/\text{M}_{\odot})< 12$ and the best fit-parameters are reported in table~\ref{fitparam}. 

Even though it is not the main focus of the present work  we also show with a blue star in figure~\ref{MSR}, for completeness, the positions in the $M_*-R_{eff}$ plane of the central dominant galaxies (CDGs). CDGs are identified in this work as the closest bright galaxy to the peak of X-ray emission. As expected, these galaxies are among the most massive and largest galaxies in the cluster. We notice that for some of them (i.e. RX0152, RCS2319, XMM1229), the automated fit delivered by \textsc{galapagos} did not converge, so we did a new fit forcing $n=4$ while keeping the values obtained with the first fit for the remaining parameters. The brightest cluster galaxies (BCGs) identified in Lidman et al. (in prep.) as the brightest galaxies in the K-band are also marked in the mass-size plane for each cluster (notice that the BCGs selected by Lidman et al. (in prep) in RCS2345 and XMM0223 have a late-type morphology).


The first result is that the slopes of the MSRs of early-types galaxies living in clusters are consistent at $1\sigma$ up to $z\sim1.2$ being the typical value $\beta=0.49\pm0.08$ which is also consistent with previous works without environment distinction \citep[e.g, ][]{newman12, cimatti12}. In the three most distant clusters of our sample we measure a smaller slope $\beta=0.27\pm0.06$, that might indicate a lack of massive and large ETGs at these higher redshifts. This difference could be due to cluster to cluster variations or a real trend at high redshift but more statistics are required to make a firm conclusion. 

{\bf We notice that the mass-size relations of XMM2235 and RCS1252  have already been studied in previous works (e.g. \citealp{2010A&A...524A..17S} et al. 2010, Rettura et al. 2010) respectively. The results are consistent within the uncertainties expected if one corrects for the different IMFs used.} 

\begin{table*}
\begin{center}
\begin{tabular}{ c c c c c }
\hline
\hline
Cluster & $z_{cl}$ & $\kappa \pm \Delta \alpha$ & $\beta \pm \Delta \beta$ & $\sigma$ \\
\hline
RXJ0152   &  0.84&  $-5.5  \pm   0.3$ &$0.54 \pm  0.09$& 0.23   \\
RCS2319   &  0.91&  $-5.1 \pm   0.5$  &$0.5 \pm  0.1$& 0.18   \\
XMMJ1229  &  0.98&  $-3.6  \pm  0.4$ & $0.4 \pm 0.1$& 0.17   \\
 RCS0220   &  1.03&  $-4.4  \pm  0.4$ & $0.4 \pm 0.1$& 0.19   \\
 RCS2345   &  1.04&  $-5.6  \pm  0.6$ & $0.6 \pm 0.2$& 0.19   \\
 XMMJ0223  &  1.22&  $-5.2  \pm  0.6$ & $0.5 \pm 0.2$& 0.23   \\
 RDCSJ1252 &  1.23&  $-3.0  \pm  0.5$ & $0.3 \pm 0.1$& 0.22   \\
 XMMU2235  &  1.39&  $-2.0  \pm  1.0$ & $0.2 \pm 0.3$& 0.25   \\
 XMMJ2215  &  1.45& $-3.0   \pm  1.0$ & $0.3 \pm 0.3$& 0.26   \\
\hline
\end{tabular}
\caption{Fit parameters of the mass-size relation for each cluster as $log(R_e/\text{kpc}) = \kappa + \beta \times log(M/\text{M}_{\odot})$.}
\label{fitparam}
\end{center}
\end{table*}


In the following sections we focus on the environmental dependence of the mass-size relation. For that purpose,  we gather all passive ETGs in clusters in three redshift bins ($0.7 \leq z < 0.9$, $0.9 \leq z < 1.1$ and $1.1 \leq z < 1.6$) in order to increase statistics and assume that the slope of the relation is constant in that redshit range. We therefore consider 2 clusters in the first bin (RX0152 and RCS2319), 3 clusters in the second bin (XMM1229, RCS0220 and RCS2345) and 4 clusters in the third bin (XMM0223, RDCS1252, XMM2235 and XMM2215).

\subsection{The mass-size relation of early-type galaxies in different environments}

\begin{figure*}
\begin{center}
\includegraphics[width = 0.65\textwidth]{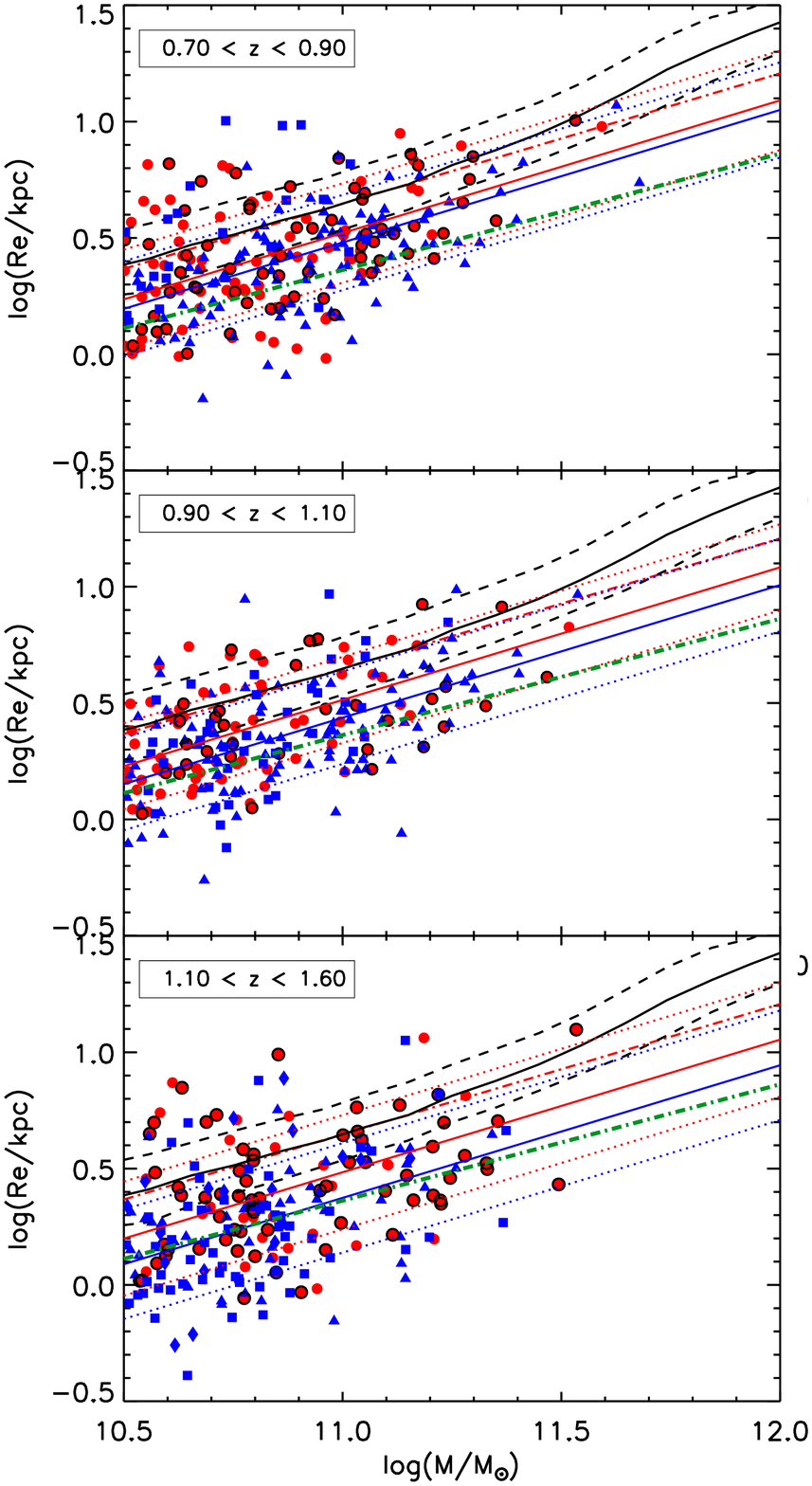}
\caption{Mass-size relation (MSR) of passive early-type galaxies in clusters and in the field. The local mass-size relation of \protect\cite{bernardi12} is represented in black lines and the red dashed-dotted line shows the Shen et al. (2003) relation. Passive ETGs in clusters are represented in red circles and passive ETGs in the field are the blue symbols. Blue squares are from the CANDELS sample of \protect\cite{newman12} and blue diamonds are the GOODS sample from \protect\cite{raichoor12}. Blue and red lines correspond respectivelly to the fit of the MSR with a fixed slope for field sample and for cluster sample and the dotted lines, the fit $\pm 1\sigma$. The green dash-dotted line corresponds to the surface density threshold : we select super-dense galaxies below that line.}
\label{MSR_field}
\end{center}
\end{figure*}

In Figure~\ref{MSR_field}, we show the mass-size relation of passive ETGs in clusters and in the field in the three different redshift bins described above ($0.7<z<0.9$, $0.9<z<1.1$ and $1.1<z<1.6$) and summarize the best-fit parameters using a power-law in table~\ref{fitfieldcl}. 


In order to look for differences in the intercepts ($\kappa$) of field and cluster mass-size relations, we fix the slope at $\beta = 0.57$ (the value measured in the local universe and compatible with our measurements) and perform a new fit. Cluster galaxies tend to present larger intercept values with the difference increasing with redshift. In fact it goes from $9\%$ at $z\sim0.8$ to $23\%$ at $z\sim1.5$. The latter is significant at more than $3\sigma$. This result suggest that cluster red-sequence ETGs are on average larger than field galaxies at fixed stellar mass. We precisely quantify this effect in section~\ref{sec:size_evol}. We emphasize that this result is not in contradiction with the fact that there is a larger fraction of SDGs in clusters since the analyzed populations are different (there is no morphological selection for SDGs). 


\begin{table*}
\begin{center}
\begin{tabular}{ c c c c c }
\hline
\hline
Redshift & Environment   & $\kappa \pm \Delta \kappa$ & $\beta \pm \Delta \beta$ & $\sigma$ \\
\hline
$[0.7, 0.9]$& Cluster    & $-5.2 \pm 0.3$& $+0.52  \pm  0.08$& 0.22  \\
            &  Field     & $-4.7 \pm 0.2$& $+0.47  \pm  0.07$& 0.19  \\
            & Cluster    & $-5.75 \pm 0.02$& $+0.57$& 0.21  \\
            &  Field     & $-5.79 \pm 0.02$& $+0.57$& 0.20  \\
\hline                                   
$[0.9, 1.1]$& Cluster    & $-4.8 \pm 0.3$& $+0.48  \pm  0.08$& 0.19  \\
            & Field      & $-5.8 \pm 0.2$& $+0.57  \pm  0.07$& 0.19  \\
            & Cluster    & $-5.75 \pm 0.02$& $+0.57$& 0.18  \\
            &  Field     & $-5.83 \pm 0.02$& $+0.57$& 0.20  \\
\hline                                   
$[1.1, 1.6]$& Cluster   & $-3.3 \pm 0.3$& $+0.34  \pm  0.10$& 0.25  \\
            & Field      & $-5.2 \pm 0.3$& $+0.50  \pm  0.10$& 0.24   \\
            & Cluster    & $-5.78 \pm 0.03$& $+0.57$& 0.24  \\
            &  Field     & $-5.89 \pm 0.02$& $+0.57$& 0.23  \\
\hline
\end{tabular}
\caption{Fit parameters of the mass-size relation for field and cluster galaxies as $log(R_e/\text{kpc}) = \kappa+ \beta \times log(M/\text{M}_{\odot})$ with free slope and fixed slope $\beta=0.57$.}
\label{fitfieldcl}
\end{center}
\end{table*}


\subsection{Size evolution of massive early-type galaxies in different environments}
\label{sec:size_evol}

We now focus on the size evolution over the $\sim2.5$~Gyr covered by our data. We will use in the following as primary size estimator the mass-normalized size ($\gamma$) as defined by \cite{newman12} and \cite{cimatti12}:
\begin{equation}
\gamma=R_e/M_{11}^{\beta}
\end{equation}
with 
\begin{equation}
M_{11} = M_*/10^{11} \; \text{M}_\odot
\end{equation}
and 
\begin{equation}
\beta=0.57
\end{equation}

By using this quantity we intentionally remove the correlation between $R_e$ and $M_*$ which could produce spurious differences in the size distributions of different samples if the mass distributions are not identical. {\bf This is basically equivalent to follow the evolution of the intercept on the mass-size relation.} The $\beta$ parameter is calibrated on the local mass-size relation and the main assumption we make is that the slope of the relation does not change significantly with redshift which, as shown in the previous section, is consistent with our sample at first level. 
We notice that our main results are robust against changes in $\beta$ of around 10~$\%$. 

\begin{figure*}
\begin{center}
\includegraphics[width = 0.98\textwidth]{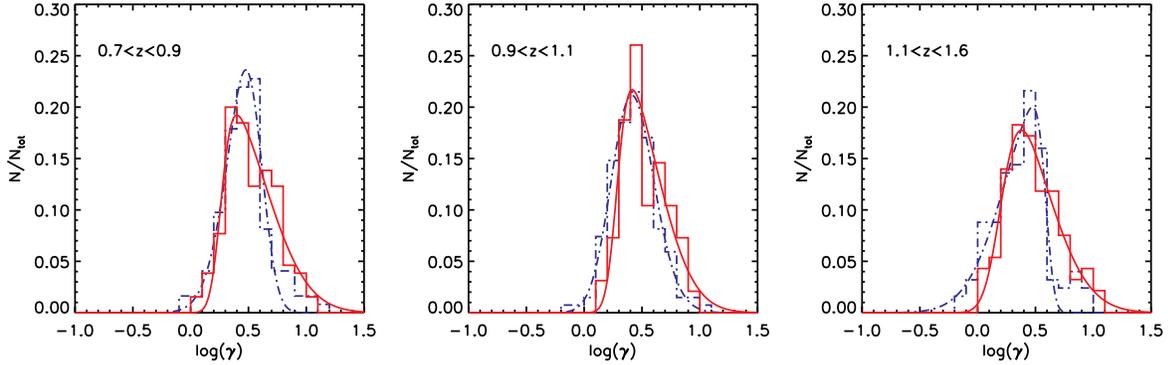}
\caption{Distributions of $log({\gamma})$ of passive early-type galaxies with $\log(M/\text{M}_{\odot}) \geq 10.5$ in the field (blue dashed line) and in clusters (red line) for 3 bins of redshifts. The red line and blue dashed line show the best fit skew normal distribution for cluster and field galaxies respectively. Galaxies in clusters show a population of galaxies with larger sizes that are not found in the field (see text for details).}
\label{SH1}
\end{center}
\end{figure*}

We first show in figure~\ref{SH1} the $\gamma$ distributions in three redshift bins, in clusters and in the field. In some redshift bin, Kolmogorov-Smirnov (KS) tests present small values (see table~\ref{KStest_fieldcl}), but never low enough ($P<0.05$), to clearly state that the two distributions are different. 


In order to estimate the mean sizes at a given redshift, the size distributions for cluster and field galaxies are fitted with a model in which $log(\gamma)$ follows a skew-normal distribution, as previously done by \cite{newman12}. The best fit models are over-plotted in figure~\ref{SH1}. 

This model has the advantage of better describing eventual asymmetries in the size distribution. The skew normal distribution has indeed three
parameters: the mean $<log(\gamma)>$, the standard deviation $\sigma_{log({\gamma})}$, and a shape parameter $s$ that is related to the skewness:

\begin{equation}
P(log({\gamma}))=\frac{1}{\omega\pi}e^{-\frac{(log({\gamma})-\psi)^2}{2\omega^2}}\int_{-\infty}^{s(\frac{log({\gamma})-\psi}{\omega})}e^{-\frac{t^2}{2}}dt
\end{equation}

We then estimate $<log({\gamma})>$ at a given redshift as the mean of the best fit skew normal distribution, which is given by $<log({\gamma})>=\psi+\omega\delta\sqrt{2/\pi}$ where $\delta=s/\sqrt{1+s^2}$. Uncertainties on sizes are then computed by bootstrapping, i.e. we repeat the computation of each value $1000$ times removing one element each time and replace it by another, and compute the error as the scatter of all the measurements.

\begin{figure}
\begin{center}
\includegraphics[scale=.8]{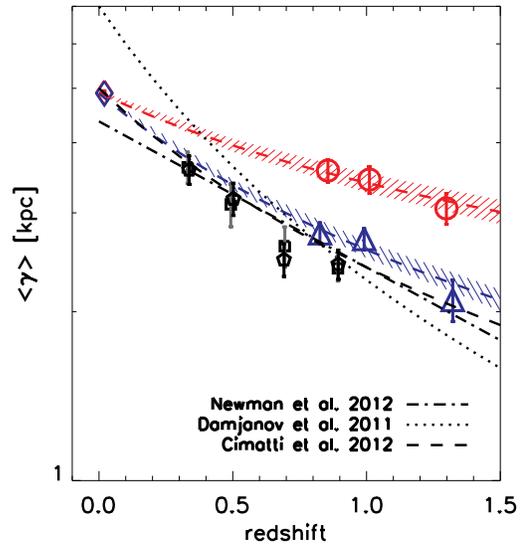}
\caption{Evolution of $<\gamma>$ for passive ETGs with $\log(M/\text{M}_{\odot}) \geq 10.5 $ in clusters (red circles) and in the field (blue triangles). Red and blue diamonds show the value of $<\gamma>$ in the local universe in clusters and in the field respectively (see text for details). Black squares and diamonds are the values in the field an in groups from Huertas-Company et al. (2013). Blue and red dashed lines show the best-fit model $\gamma\propto(1+z)^\alpha$ for field and cluster galaxies respectively. Cluster passive ETGs are on average larger at $z\sim1$ and present a less steep evolution than field galaxies at fixed stellar mass.}
\label{evol_fieldcl}
\end{center}
\end{figure}

Figure~\ref{evol_fieldcl} shows now the redshift evolution of $<log({\gamma})>$ for passive ETGs in clusters and in the field with stellar masses above $3 \times 10^{10} \; \text{M}_{\odot}$. Cluster ETGs have mean values of $<log({\gamma})>$ $\sim1.5$ times larger than field ETGs of the same stellar mass (see also table~\ref{tbl:gamma}). We notice that the position  of the peaks of the distributions of clusters and field galaxies in figure~\ref{SH1} are consistent so we would have not found the same results when considering a symmetric gaussian fit. In fact the difference is explained because the distribution of cluster galaxies is more skewed towards larger values of $\gamma$ as seen in figure~\ref{SH1}, e.g. cluster ETG show a population skewed towards larger galaxies with respect to the field. In particular, the skewness of the best model fit  for cluster galaxies is clearly positive ($s\sim0.8$ for the three redshift bins) while the distribution of field galaxies present a negative skewness ($-0.$ to $-0.4$). As a sanity check, we over plot the size evolutions recently reported by \cite{newman12, cimatti12, damjanov11} using independent datasets without environment distinction. Our results in the field are globally consistent with previous measurements by \cite{newman12} and \cite{cimatti12}. The discrepancy is slightly larger ($\sim2\sigma$) with \cite{damjanov11} who measure $\gamma \propto (1+z)^{-1.62 \pm 0.34}$ and a larger zero point. We also show in figure~\ref{evol_fieldcl} the sizes of ETGs in groups and in the field from \cite{huertas12}, which also globally lie on the same relation. The points at  $z\sim0$ are computed from the SDSS by cross-correlating the group catalog of Yang et al. (2007) updated to the DR7 and the morphological classification of Huertas-Company et al. (2011). We select ETGs in the same stellar mass range ($M_*>3\times10^{10}$) than the high-redshift sample and compute $<\gamma>$ with the same methodology. We select as cluster galaxies those living in the most massive haloes ($M_h/M_\odot>10^{14}$) and field galaxies are selected in the low mass end of the halo mass function i.e. $M_h/M_\odot<10^{13}$ (see Huertas-Company et al. 2013 for more details). 

Since the difference between cluster and field galaxies is not seen in the local universe,  field ETGs follow $\gamma \propto (1+z)^{\alpha}$ with $\alpha = -0.92 \pm 0.04$ and cluster ETGs have a value of $\alpha = -0.53 \pm 0.04$ (see table~\ref{fitevol}). The evolutions are therefore different at more than $3\sigma$. 

\begin{figure}
\begin{center}
\includegraphics[scale=0.8]{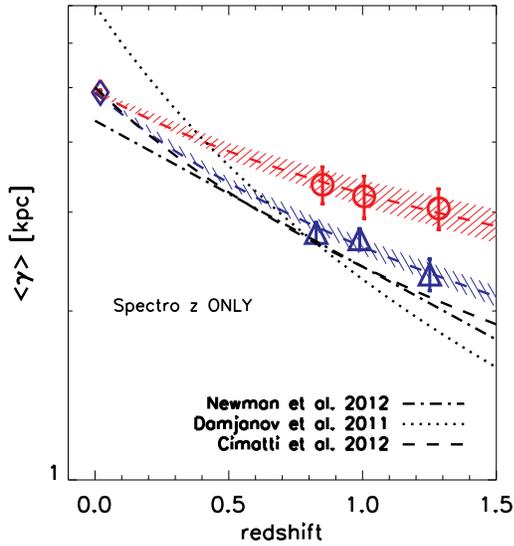}
\caption{Same as figure~\ref{evol_fieldcl} but only spectroscopically confirmed members are included in the cluster sample.}
\label{fig:specz}
\end{center}
\end{figure}

The result is robust to morphological classifications. If we consider cluster and field ETGs visually classified, the small differences in the morphological classifications reported in table~\ref{Nsample} and section~\ref{sec:morpho} do not change the trends on the size evolution, i.e. the $\alpha$ values from the best fits remain the same within the error bars. Also, when we consider only spectroscopically confirmed members, the result remains unchanged (fig.~\ref{fig:specz}). Finally, since the COSMOS sample is shallower than the other samples (see section~\ref{sec:field}), incompleteness might have an impact in the size evolution. We have thus checked that our results do not change when the COSMOS sample is removed,  even if our uncertainties become larger due to lower statistics.

\begin{table*}
\begin{center}
\begin{tabular}{ c c c }
\hline
\hline
redshift bin &  Clusters &  Field  \\
\hline
$[0.7, 0.9]$ &   $3.55\pm0.16$ &  $2.76\pm0.12$\\
$[0.9, 1.1]$ &  $3.43\pm0.18$ & $2.69\pm0.11$ \\
$[1.1, 1.6]$ &  $3.05\pm0.19$  &  $2.09\pm0.17$\\
\hline
\end{tabular}
\caption{Mean values of $log(\gamma)$ measured in our sample} 
\label{tbl:gamma}
\end{center}
\end{table*}

\begin{table*}
\begin{center}
\begin{tabular}{ c c c c c c }
\hline
\hline
Mass & Redshift & N$_{\text{cluster}}$   & N$_{\text{field}}$  & K-S &  Kuiper  \\
\hline
$log(M/\text{M}_{\odot}) \geq 10.5 $ & $0.7 \leq z < 0.9$ &   130 & 123  & 0.05 & 0.14  \\
                                     & $0.9 \leq z < 1.1$ &   96  & 135  & 0.06 & 0.25  \\
                                     & $1.1 \leq z < 1.6$ &   94  & 125  & 0.03	 & 0.29   \\ 
\hline
\end{tabular}
\caption{ Results of Kolmogorov-Smirnoff and Kuiper statistical tests applied to field and cluster ETGs mass-normalized radius distributions for different redshift bins. N$_{\text{cluster}}$ and N$_{\text{field}}$ indicate respectivelly the number of cluster galaxies and the number of field galaxies in each sample.}
\label{KStest_fieldcl}
\end{center}
\end{table*}

\begin{table*}
\begin{center}
\begin{tabular}{c c c c }
\hline
\hline
Sample & Mass & $\alpha \pm \Delta\alpha$ & $\beta \pm \Delta\beta$  \\
\hline
Cluster ETGs & $log(M/\text{M}_{\odot}) \geq10.5$     & $-0.53 \pm 0.04$  & $4.91 \pm 0.04$\\
\hline
Field ETGs  & $log(M/\text{M}_{\odot}) \geq 10.5 $  & $-0.92 \pm 0.04$  & $4.89 \pm 0.02$ \\ 
\hline
\end{tabular}
\caption{Fit parameters of the size evolution of cluster and field ETGs as $\gamma = \beta \times (1+z)^{\alpha}$.}
\label{fitevol}
\end{center}
\end{table*}

\section{Discussion}
\label{sec:disc}

Massive ETGs living in massive clusters ($M/M_\odot>10^{14}$) at $0.8<z<1.5$ appear to be on average $1.5$ times larger than galaxies of the same stellar mass residing in the field. Similar results have been obtained by Papovich et al. (2012) on a single cluster at $z\sim1.6$ but no significant difference was measured by Raichoor et al. (2012) in the Lynx supercluster at $z\sim1.3$.

Interestingly this size difference is not seen in the nearby Universe (Huertas-Company et al. 2013, Poggianti et al. 2013) where cluster and field galaxies present similar sizes. If cluster galaxies are growing faster at earlier epochs they should be somehow caught up by the galaxies living in the less dense environments to end up with the same size distributions. In figure~\ref{fig:SDSS} we compare the distributions of $log(\gamma)$ in the three redshift bins probed by our sample with the same distribution in the SDSS.

Clearly the local distributions for cluster and field galaxies are undistinguishable, peak at larger sizes and are symmetric (e.g. Huertas-Company et al. 2013). The smallest galaxies have disappeared  in the low redshift sample. The small end of the high redshift distribution gradually fills up the peak of the local : there is a transition of the peak of the galaxy distribution from smaller to larger sizes. During this transition phase, the high redshift size distributions are skewed towards larger values and, according to our data, this process has already started in clusters at $z\sim1$, while it is not yet observed in the field. 

\begin{figure*}
\begin{center}
\includegraphics[scale=0.8]{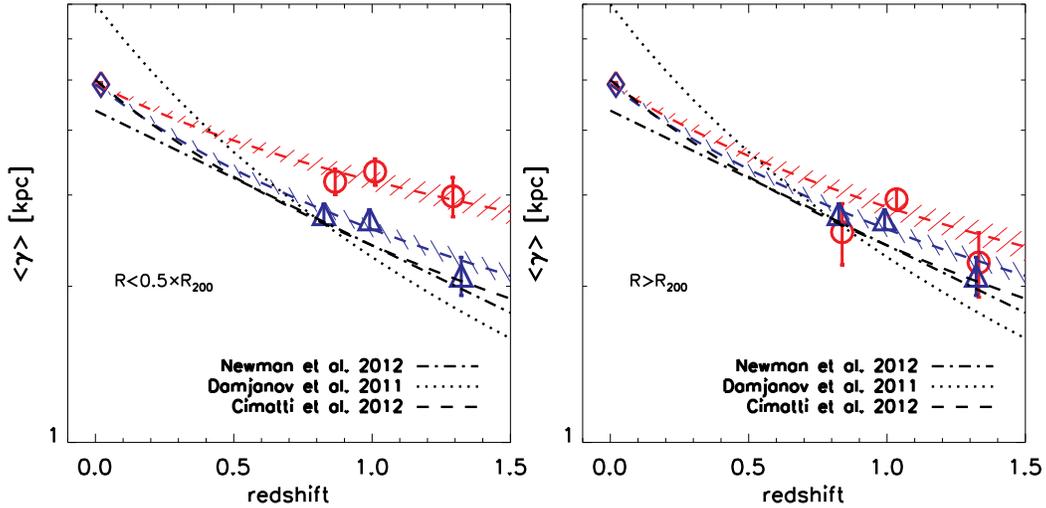}
\caption{Redshift evolution of $\gamma$ in clusters (red circles) and in the field (blue triangles) for galaxies living in the central parts of the cluster ($R<0.5*R_{200}$, left panel) and in the outskirts ($R>R_{200}$, right panel) as labelled. Symbols are the same than for figure~\ref{evol_fieldcl}. The size difference between cluster and field galaxies is mainly driven by the galaxies living in the central parts of the clusters.}
\label{fig:outskirts}
\end{center}
\end{figure*}

\begin{figure*}
\begin{center}
\includegraphics[scale=0.6]{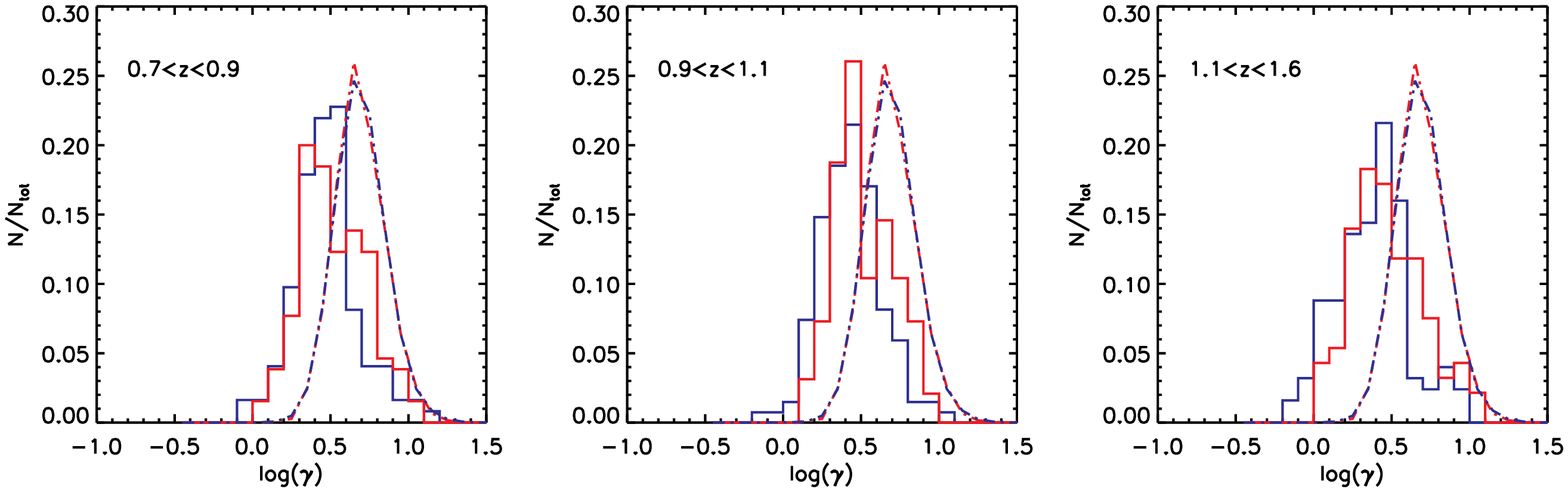}
\caption{Distributions of $log({\gamma})$ in three different redshift bins as labelled, for cluster and field galaxies in our sample (red and blue solid lines respectively) as compared to the distribution of galaxies at $z\sim0$ from the SDSS in clusters and in the field (dashed-dotted red and blue lines respectively).  }
\label{fig:SDSS}
\end{center}
\end{figure*}

If at least part of the evolution seen in $\gamma$ is due to mergers, the difference we see between cluster and field galaxies might reflect the fact that on average cluster galaxies at $z\sim1$ have experienced more mergers than field galaxies at the same epoch, probably during the formation phase of the clusters when velocity dispersions are lower. Since the size evolution from $z\sim1$ to present is then slower in clusters than in the field (as shown in section~\ref{sec:size_evol}),  the mechanism that increased sizes of cluster galaxies should drop its efficiency, e.g. cluster galaxy merger rates become lower than in the field due probably to the increase of the galaxy velocity dispersion. If this is true, we should observe that these larger galaxies in high redshift clusters are concentrated in their cores, where in these early epochs dynamical friction and higher densities cause a higher merger rate with respect to their outskirts. In figure~\ref{fig:outskirts}, we show that in fact the galaxies which are mainly driving the size difference  observed in this work are those living in the central parts of the clusters. Galaxies in the cluster outskirts, which are entering the cluster potential at the epoch of observation ($R>R_{200}$) indeed have similar sizes than galaxies in the field at the same epoch. This supports the idea that galaxies in the core have been processed at earlier epochs. On the other hand, in the field, the merger rates are known to show little evolution from $z\sim1$ to present (e.g \citealp{2011ApJ...742..103L}),  which would explain why the field galaxies would evolve faster than in the clusters. Studying a sample of cluster/groups at even earlier epochs $z\sim2$ should help in answering these questions.

\begin{figure*}
\begin{center}
\includegraphics[width=0.98\textwidth]{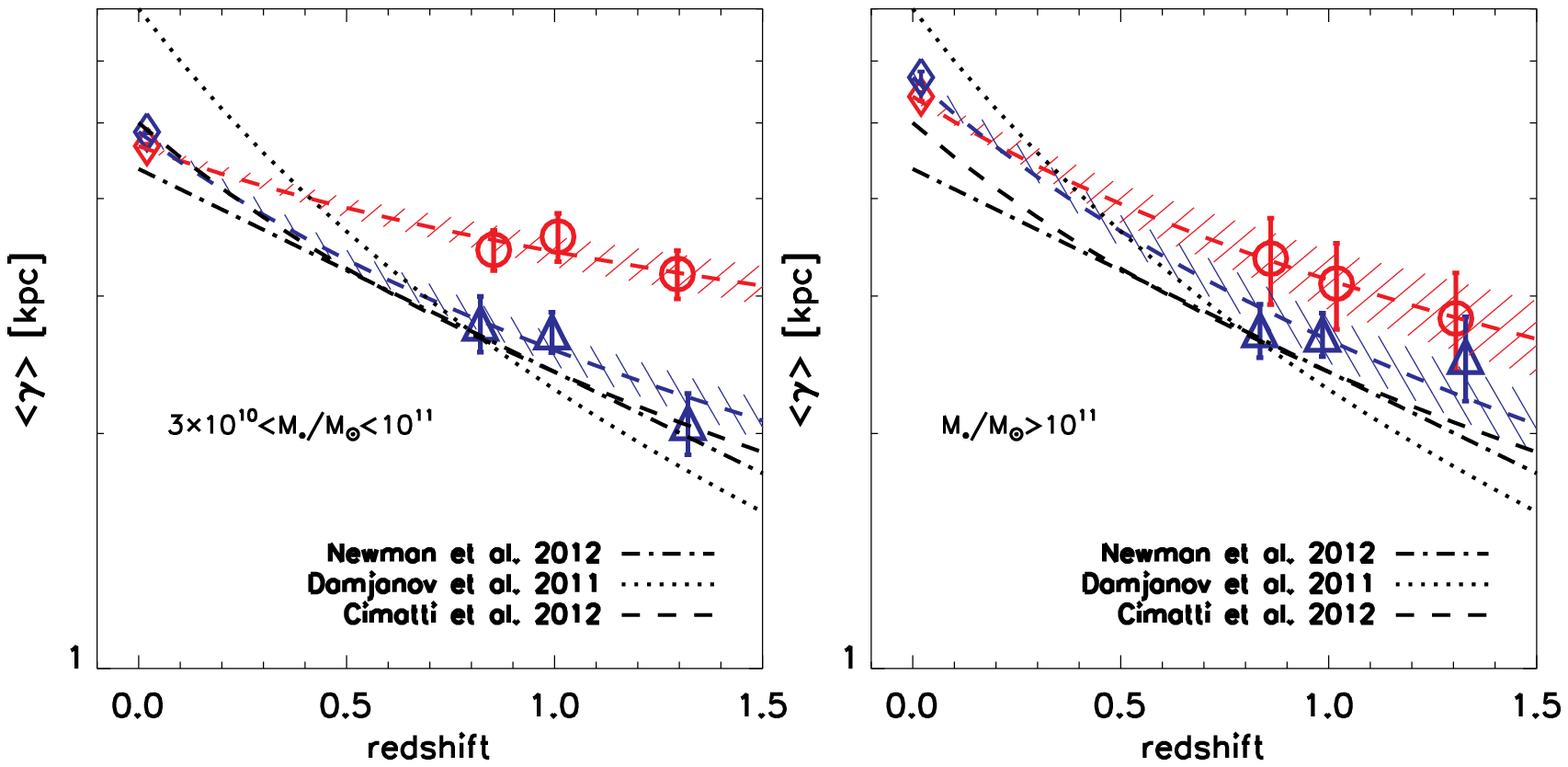}
\caption{Redshift evolution of $\gamma$  in clusters (red circles) and in the field (blue triangles) for galaxies in the stellar mass-range $3\times10^{10}<M_*/M_\odot<10^{11}$ (left panel) and $M_*/M_\odot>10^{11}$ (right panel). Symbols are the same than for figure~\ref{evol_fieldcl}.}
\label{fig:mass_range}
\end{center}
\end{figure*}

To better understand which galaxies are increasing their sizes in clusters, in fig.~\ref{fig:mass_range} we split our sample in two bins of stellar mass ($10.5<log(M_*/M_\odot)<11$ and $log(M_*/M_\odot)>11$).  This threshold is selected since $10^{11} M_*/M_\odot$ appears to be a critical mass above which galaxy evolution is expected to be dominated by mergers \citep[e.g.][and references therein]{delucia06, khochfar11, shankar11} so we might naturally expect that the behavior against environment could differ for these two populations. Interestingly, we find that for the most massive galaxies ($M_*/M_\odot>10^{11}$) the difference is somehow less pronounced. This might be evidence that very massive galaxies in both environments have experienced similar size growth, even though we have lower statistics in that bin to establish a clear conclusion. 

Galaxies with mass $10.5<log(M_*/M_\odot)<11$ show larger sizes in clusters. In this mass range, mergers should not be as efficient as for galaxies of higher masses. However, the environmental differences that we observe are driven by these masses, suggesting that other mechanisms might contribute to the size enlargement. In fact, an important point is that the evolution of $\gamma$ does not measure the individual evolution of galaxies. As several works have pointed out (e.g. \citealp{2009ApJ...698.1232V}, Newman et al. 2012, \citealp{2013arXiv1302.5115C}) the evolution we see could be partially or even dominated by the quenching of new galaxies which enter the selection at later epochs. In this case, our results would reflect the fact the quenching is more efficient in the cluster environment, so it happens at earlier epochs. This would explain also why the environmental size differences are larger in the lower mass bin, since they are quenched later than the massive and more efficiently in clusters (e.g., Thomas et al. 2005).


It is  unclear how and why we reach a perfect match between the size distributions in the two environments at $z\sim 0$, and what happens at $z<1$ . Huertas-Company et al. 2013 have shown that in the COSMOS field, there are not significant differences in the mass-size relation and size evolution of galaxies in groups and in the field. However, for a DEEP2 field spectroscopic sample,  \cite{cooper12} found that larger galaxies with high Sersic index ($n>2.5$) preferentially live in dense environments (defined as the number of neighbors).

\section{Summary and conclusions}

We have studied the mass-size relations as well as the size evolution of 319 passive ETGs with $\log(M/\text{M}_{\odot}) > 10.5$ living in nine well-known rich clusters between $z=0.8$ and $z \sim 1.5$. The sample is 80\% complete. This is the largest sample of cluster galaxies at those redshifts used for that kind of study. The results are compared with the ones obtained on a sample of 382 field ETGs in the same mass and redshift range.  \\

Our main results are summarized in the following:

\begin{itemize}
	\item The slopes of the mass-size relations of ETGs in clusters do not change significantly up to $z\sim1.2$ being the typical value $\beta=0.49\pm0.08$ which is also consistent with previous works at lower redshifts without environment distinction. Our results are in favor of a very mild evolution of the slope of the MSR of early-type galaxies from $z\sim1.2$ independently of the environment. For the three clusters at $z>1.2$,  we measure $\beta=0.27\pm0.06$, that might indicate a lack of massive and large ETGs at these higher redshifts.
	

	\item The zeropoint of the MSR changes with time. Cluster ETGs with $log(M_*/M_\odot>10.5)$ roughly doubled their median size from $z\sim1.5$. Our results are in agreement with previous published results without environment distinction.

	\item Cluster red-sequence ETGs have mass normalized sizes  $\gamma$ $\sim1.5$ times larger than galaxies in the field and show a less steep evolution with evolution coefficients  ${-0.53\pm0.04}$, ${-0.92\pm0.04}$ for clusters and field, respectively.

\end{itemize}

In future work we will carefully confront our observational results with detailed model predictions (Shankar et al., in prep.).
\vspace{0.5cm}

\textit{ The authors would like to thank G. Rousset for pointing out interesting suggestions and E. Daddi, V. Strazzulo, R. Gobat for stimulating discussions. RD gratefully acknowledges the support provided by the BASAL Center for Astrophysics and Associated Technologies (CATA), and by FONDECYT grant No. 1130528. }


\appendix

\section{Cluster sample}
\label{sec:append_cl}
\begin{itemize}

	\item RX~J0152-1357 (hereafter, RX0152) at $z=0.84$ was discovered in the ROSAT Deep Cluster Survey \citep[RDCS;][]{rosati98,dellaceca00} as an extended double core X-ray source. \cite{demarco05, demarco10} confirmed spectroscopically 134 galaxies as cluster members. The velocity dispersion of the  the most massive of the central sub-clusters is $\sigma \sim 920$ km/s \citep{demarco05}. Its virial mass derived from the X-ray measurement is $M_{200} = 7.3^{+1.8}_{-1.7} \times 10^{14} \; \text{M}_{\odot}$ \citep{ettori09} whereas its mass derived from weak-lensing analysis is $M_{200} = 4.4^{+0.7}_{-0.5} \times 10^{14} \; \text{M}_{\odot}$ \citep{jee11}. This cluster was observed with ACS WFC in November 2002, in the $F625W$ ($r_{625}$), $F775W$ ($i_{775}$) and $F850LP$ ($z_{850}$) bandpasses. The exposure time was of $19000$~s, $19200$~s and $19000$~s, respectively. Our NIR image in the Ks-band was acquired using HAWK-I \citep{pirard04, casali06} on Yepun (UT4) on the Very Large Telescope (VLT) at the ESO Cerro Paranal Observatory in October 2009 and has a PSF width of $0.4"$.\\

\item{RCS 2319+0038 (hereafter, RCS2319) at $z=0.91$}:
The clusters with the RCS prefix were observed in the context of the Red-sequence Cluster Survey (RCS) \citep{gladders05}. The virial mass derived from the X-ray measurements of \cite{hicks08} is $M_{200} = 5.4^{+1.2}_{-1.0} \times 10^{14} \; M_{\odot}$. RCS2319 was observed with ACS WFC in the $i_{775}$ and $z_{850}$ bandpasses in May 2006 with a total exposure time of $2400$~s and $6800$~s, respectively. The NIR images were acquired in the Js-band using ISAAC on Melipal (VLT-UT3; \cite{morwood98b}) in July 2003 with an average PSF width of $0.63"$ and  in the Ks-band using HAWK-I in November 2009, with a PSF width of $0.47"$. We have 11 spectroscopic confirmed members \citep{gilbank08, gilbank11, meyers10}.

\item{XMMU~J1229+0151 (hereafter, XMM1229) at $z=0.98$} was initially detected in the XMM-Newton Distant Cluster Project \citep[XDCP, ][]{bohringer07, fassbender07}. This clusters is a rich, hot and X-ray luminous galaxy cluster \citep{santos09}. The mass measured with lensing is $M_{200} = 5.3^{+1.7}_{-1.2} \times 10^{14} \; M_{\odot}$ \citep{jee11} whereas the virial mass from X-ray measurements is $M_{200} = 5.7^{+1.0}_{-0.8} \times 10^{14} \; M_{\odot}$ \citep{santos09}. 27 cluster members were spectroscopically confirmed with the VLT/FORS2 spectrograph \citep{santos09}. In the framework of the Supernova Cosmology Project \citep{dawson09}, we obtained ACS/WFC images  in the $i_{775}$ and the $z_{850}$ bandpasses  in December 2005, for total exposures of $4110$~s and $10940$~s, respectively. NIR imaging in the J-band was acquired using SOFI \citep{morwood98a} at the New Technology Telescope (NTT) at the ESO/La Silla observatory in March 2007, whereas the Ks-band imaging was acquired using HAWK-I in January 2010. The J-band data have a PSF width of $0.98 "$ and the K-band have a PSF width of $0.41"$. This cluster was also observed in the $F160W$ bandpass with the WFC3 on HST in May 2010, with a PSF width of $0.3"$, and a pixel scale of $0.1282$ \arcsec/pixel.

\item{RCS 0220-0333 (hereafter, RCS0220) at $z=1.03$} is an optically rich cluster at $z=1.03$ with 14 spectroscopic confirmed members \citep{meyers10, gilbank11}. The weak-lensing mass of the cluster is $M_{200} = 4.8^{+1.8}_{1.3} \times 10^{14} \; M_{\odot}$ with a predicted velocity dispersion of $881^{+68}_{-74} \; \text{km} \; \text{s}^{-1}$ \citep{jee11}. This cluster was observed with ACS/WFC in the $i_{775}$ and $z_{850}$ bandpasses in 2005 with a total exposure of $2955$~s and $14420$~s, respectively. The NIR images were acquired in the Js-band using ISAAC in October 2002, with a PSF width of $0.47"$  and  in the Ks-band using HAWK-I in January 2010, with a PSF width of  $0.35"$.

\item{RCS 2345-3633 at $z=1.04$}
RCS~2345-3633 is an optically rich cluster at $z=1.04$ with 23 spectroscopic confirmed cluster members \citep{meyers10, gilbank11}. The virial mass estimated by weak-lensing in \cite{jee11} is $M_{200} = 2.14^{+1.1}_{-0.7} \times 10^{14} \; M_{\odot}$. As the previous one, this cluster was observed with ACS/WFC in the $i_{775}$ and $z_{850}$ bandpasses in July 2006 with a total exposure of $4450$~s and $9680$~s, respectively. The NIR images were acquired in Js-band using ISAAC in July 2003 and  in Ks-band using HAWK-I in October 2010. The PSF width for HAWK-I image is $0.39"$ and for ISAAC image, it is $0.56"$.

\item{XMMLSS~0223-0436 (hereafter, XMM0223) at $z=1.22$}
	XMMLSS~0223-0436 was discovered in the XMM Large Scale Survey (LSS) \citep{pierre04, andreon05}. \cite{jee11} estimated that the virial mass of this cluster from weak-lensing analysis is $M_{200} = 7.4^{+2.5}_{-1.8} \times 10^{14} \; M_{\odot}$, more than 2 times larger than the virial mass from X-ray measurements\citep{bremer06}. We used optical images from ACS WFC in $F775W$ and $F850LP$ bandpasses acquired in September and July 2005 with a total exposure time of $3380$~s and $14020$~s, respectively. XMMLSS~0223 has NIR imaging in the Js- and Ks-band obtained with HAWK-I in November 2009. The NIR PSF width is of $0.40"$ in the Js-band and of $0.38"$ in the Ks-band. 23 cluster members were spectroscopically confirmed \citep{bremer06,meyers10}.\\

\item{RDCS J1252-2927 (hereafter, RDCS1252) at $z=1.23$} was discovered  in the ROSAT Deep Cluster Survey \citep{rosati98} and confirmed as a cluster at $z=1.23$ based on an extensive spectroscopic campaign using the VLT \citep{rosati04,lidman04}. The viral mass based on a lensing analysis on deeper ACS images is $M_{200} = 6.8^{+1.2}_{-1.0} \times 10^{14} \; M_{\odot}$ \citep{jee11} whereas X-ray measurements gives $M_{200} = 7.6 \pm 1.2 \times 10^{14} \; M_{\odot}$ \citep{ettori09}. For this cluster, we have 38 spectroscopic confirmed members from \cite{demarco07}. Imaging ACS WFC in the $i_{775}$ and $z_{850}$ bandpasses were acquired in May 2002 with exposure times of $29945$~s and $57070$~s, respectively. NIR data were obtained from ISAAC in Js- and Ks-band with a PSF width of $0.51"$ and $0.42"$ respectivelly. \\

\item{XMMU J2235-2557 (hereafter, XMM2235) at $z=1.39$}
	XMMU~J2235-2557 is one of the most massive X-ray luminous cluster at $z > 1$ with a virial mass $M_{200} \sim 6 \times 10^{14} \; M_{\odot}$ derived by X-ray measurement \citep{rosati09}. The mass from weak-lensing analysis is $M_{200} = 7.3^{+1.7}_{-1.4} \times 10^{14} \; M_{\odot}$ \citep{jee11}. Optical images were acquired using ACS WFC in the $i_{775}$ and $z_{850}$  bandpasses in June 2005. The total exposure time are $8150$~s and $14400$~s respectively. NIR imaging in Js- and Ks-band were taken using HAWK-I in October 2007 with a PSF width of $0.52"$ and $0.37"$, respectively. 31 cluster members were spectroscopically confirmed \citep{rosati09}.\\

\item{XMMXCS J2215-1738 (hereafter, XMM2215) at $z=1.45$} was the highest redshift cluster spectroscopically confirmed \citep{stanford06} until the recent discovery of ClG J0218-0510 at $z=1.62$ \citep{papovich10, tanaka10}. The virial mass from X-ray measurement is $M_{200} = 2.0^{+0.5}_{-0.6} \times 10^{14} \; M_{\odot}$ \citep{hilton10} and the one derived from weak-lensing is $M_{200} = 4.3^{+3.0}_{-1.7} \times 10^{14} \; M_{\odot}$ \citep{jee11}. Optical images were acquired using ACS WFC in the $i_{775}$ and $z_{850}$  bandpasses in April 2006. The total exposure times are $3320$~s and $16935$~s, respectively. NIR imaging in Js- and Ks-band were taken using HAWK-I in September and October 2009 with a PSF width of $0.54"$ and $0.43"$ respectively. 52 cluster members were spectroscopically confirmed by \citep{hilton10}.

\end{itemize}

\bibliographystyle{aa}
\bibliography{bibfile}

\label{lastpage}

\end{document}